\documentclass[manuscript,screen]{acmart}
\usepackage{balance}
\usepackage{natbib}
\usepackage{amsmath}
\usepackage{algorithm}
\usepackage{graphicx}
\usepackage{xcolor}
\usepackage{array}
\usepackage[caption=false,font=normalsize,labelfont=sf,textfont=sf]{subfig}
\usepackage{textcomp}
\usepackage{stfloats}
\usepackage{url}
\usepackage{verbatim}
\usepackage{texdef2015}
\usepackage{booktabs}
\usepackage[noend]{algpseudocode}

\usepackage{multirow}
\usepackage{cancel}
\usepackage{enumitem}
\usepackage{wrapfig}
\makeatletter
\newcommand*{\addFileDependency}[1]{
  \typeout{(#1)}
  \@addtofilelist{#1}
  \IfFileExists{#1}{}{\typeout{No file #1.}}
}

\AtBeginDocument{%
  \providecommand\BibTeX{{%
    Bib\TeX}}}

\def\BibTeX{{\rm B\kern-.05em{\sc i\kern-.025em b}\kern-.08em
    T\kern-.1667em\lower.7ex\hbox{E}\kern-.125emX}}

\begin{document}

\title{Learning To Communicate Over An Unknown Shared Network}

\author{Shivangi Agarwal}
\email{shivangia@iiitd.ac.in}
\orcid{0000-0002-0240-9454}
\author{Adi Asija}
\email{adiasija@gmail.com}
\author{Sanjit K. Kaul}
\email{skkaul@iiitd.ac.in}
\author{Arani Bhattacharya}
\email{arani@iiitd.ac.in}
\author{Saket Anand}
\email{anands@iiitd.ac.in}
\affiliation{%
  \institution{IIIT-Delhi}
  \city{Delhi}
  \country{India}
}



\renewcommand{\shortauthors}{Agarwal et al.}
\begin{abstract}

As robots (edge-devices, agents) find uses in an increasing number of settings and edge-cloud resources become pervasive, wireless networks will often be shared by flows of data traffic that result from communication between agents and their corresponding edge-cloud nodes (cloud compute or data resource accessed by an agent). In such a setting, any agent communicating with the edge-cloud is unaware of the state of the network resource, which evolves in response to not just the agent’s own communication at any given time but also to communication by the other agents, which stays unknown to the agent. 

We address the challenge of an agent learning a policy that allows it to decide whether or not to communicate with its cloud node, using limited feedback it obtains from its own attempts to communicate, with the goal of optimizing its utility. The policy must generalize well to any number of other agents sharing the network and must not be trained for any particular network configuration. Our proposed policy is a deep reinforcement learning model Query Net (QNet) that we train using a proposed simulation-to-real framework. Our simulation model has just one parameter and is agnostic to specific configurations of any wireless network. It however allows training an agent’s policy over a wide range of outcomes that an agent’s communication with its edge-cloud node may face when using a shared network, by suitably randomizing the simulation parameter. We propose a learning algorithm that addresses the challenges we observe in training QNet. We validate our simulation-to-real driven approach through experiments conducted on real wireless networks including WiFi and cellular. We compare QNet with other policies to demonstrate its efficacy. Our WiFi experiments involved as few as five agents, resulting in barely any contention for the network, to as many as fifty agents, resulting in severe contention. The cellular experiments spanned a broad range of network conditions, with baseline network round-trip-times ranging from a low of $0.07$ second to a high of $0.83$ second.
\end{abstract}

\ccsdesc[500]{Computing methodologies~Multi-agent reinforcement learning}
\ccsdesc[300]{Computing methodologies~Simulation types and techniques}
\ccsdesc[100]{Computer systems organization~Embedded and cyber-physical systems}
\ccsdesc[300]{Computing methodologies~Markov decision processes}
\ccsdesc[300]{Networks~Wireless access networks}

\keywords{Multi-Agent Reinforcement Learning, Scalable Multi-Agent Communication, Sim-to-Real, Domain Randomization, Wireless Networks, Edge-Cloud.}


\maketitle

\section{Introduction}
Robots that make real-time autonomous decisions are expected to impact verticals including agriculture, health care, transportation, and manufacturing. Given sensing and compute constraints, such decision making often requires compute and measurements not locally available at the robot~\cite{Liu_IEEE2019}. An edge-cloud can provide such compute and can act as a store for measurements received from other sources that may be of interest to the robot. The edge-cloud is typically a fast (a large bits per second) wireless link away from the robot. However, a fast wireless link may not guarantee low-latency as latency is significantly impacted by others sharing the wireless network with the robot, for example other robots using the network for their communication needs. It is key for a robot to adapt its communication with the edge-cloud to the current wireless network performance, as perceived by it, and given its utility.

In our work, we model an interaction of a robot (more generally an edge-device/ agent) with the edge-cloud as a message exchange that begins with a query message sent by the agent over a wireless network. The edge-cloud responds to the query message by sending back a message that contains the information desired by the agent. In general, the query message could contain measurements made by the agent with a query regarding desired processed information. Or it could be a query about a desired measurement that is not available at the agent.

Given that the wireless network is shared, sending a query message at every decision instant, agnostic to network conditions, may result in the agent experience large delays in responses from the edge-cloud that are detrimental to its utility. The challenge is for the agent to choose whether to query or not at any decision instant in a manner that adapts to the network conditions and optimizes the robot's utility. It is key to note that the state of the wireless network is unknown to an agent. The state includes the queues of packets awaiting transmission at other users sharing the wireless network and packets that are actively attempting transmission or being transmitted. Such distributed information is infeasible to obtain. The agent, however, can perceive the condition of the wireless network via the results of it querying the edge-cloud. For example, the delays in responses from the edge-cloud and the rate of response are indicative of the utilization of the network by all those sharing it. 

In our work, the decision to query is made by an agent using our proposed deep reinforcement learning (RL) model Query Net (QNet). QNet must learn a policy that chooses whether to query or not on the basis of imperfect information about the shared network.

There exists a wide variety of wireless network technologies, for example, cellular networks including variants $3$G, $4$G, and $5$G, WiFi networks $802.11a/b/g$, $802.11/n/ac$, and Bluetooth networks. These technologies differ in how they share the available wireless network resource (bandwidth) between one or more users and the maximum bit rates they offer to a user. This together with a wide range of possible demands from a variable number of users sharing a network, make it impractical to have an agent learn a policy over a representative set of experiences in deployments of real wireless networks. Also, obtaining the ground truth, which is required when an agent is learning a policy, may not be feasible in real networks. One could use high-fidelity simulations of particular technologies and for a chosen technology generate agent experiences over different network configurations. However, such simulations are compute intensive and very time consuming~\cite{adday_2024_simulation_tools, kritsis_2018_tutorial}. As with network technologies, there are many providers of edge-cloud services.

We refrain from making specific choices of network technologies and their configurations and that of edge-cloud services. Instead, we propose a low-cost low-fidelity simulation model of the network and the edge-cloud that has just a single parameter $q$ (see Figure~\ref{fig:neworkSim}), which is a probability, to generate experiences to train QNet.

A model trained using simulation must perform well in different real-world deployments. This is a significant challenge~\cite{cao-2022-cloud_edge_training, sim2real_dynamic_randomization, zero_shot_randomisation_2, tiboni-2023-recent, sim2real_object_avoidance, horvath-2022-real_data, gielis-2022-springer_book, prorok-k2021-arxiv, xie-2020-pmlr,kadian2020sim2real_acm_blog}. One must address the disparity between the simulated world in which models are trained and the real-world domains in which the trained models are deployed, termed as the ``reality gap"~\cite{cao-2022-cloud_edge_training, sim2real_dynamic_randomization, zero_shot_randomisation_2, tiboni-2023-recent, sim2real_object_avoidance}. We apply domain randomization~\cite{sim2real_dynamic_randomization}, which involves varying the parameters of a simulated environment during the training process to create diverse simulated scenarios, with the hope of covering a spectrum of real-world situations. Domain randomization comes with challenges, including choosing an appropriate simulation model with parameters to randomize~\cite{domain_random_challenge}. 

We bridge the gap between our choice of imperfect low fidelity network simulations and the real world wireless networks by training QNet over episodes generated by our simulation model by choosing the parameter $q$ over a wide range of $(0.05,1)$. We demonstrate the efficacy of using domain randomization together with our proposed simulation model, by testing QNet, trained using only simulations, in real WiFi and cellular networks,  showing that the model transfers well, without any additional training using real networks. We also compare QNet with other policies and show that QNet adapts effectively to a wide range of network conditions.

In summary, our contributions are \textbf{(1)} A deep RL model QNet that at every decision instant chooses whether or not an agent must query the edge-cloud. The decision is made using the agent's imperfect view of the network condition based on its own experiences communicating over the network. We also propose a learning algorithm for training QNet over a very wide range of network conditions. \textbf{(2)} A simulation-to-real pipeline that includes a low-cost low-fidelity single parameter simulation model that enables training QNet over a wide range of network conditions using domain randomization. \textbf{(3)} Using extensive experimentation over real wireless networks (WiFi and cellular), we demonstrate the effectiveness of the transfer of the simulation trained QNet to real networks. Further, we demonstrate the efficacy of QNet in adapting to variable network conditions by comparing its performance over real network configurations with other policies.

The rest of the paper is organized as follows. Section~\ref{sec:probSetting} details the problem formulation. Section~\ref{sec:networkAndCloudSimulationModel} describes our proposed simulation model for the network and edge-cloud. This is followed by details on the QNet model, the learning algorithm, and related challenges, in Section~\ref{sec:drlModelQNet}. In Sections~\ref{sec:experiments} and~\ref{sec:cellular_experiments} we detail our experiments on real-world WiFi and cellular networks, which demonstrate the efficacy of our proposed simulation-to-real transfer. We also compare QNet with other policies. We conclude in Section~\ref{sec:conclusions}. 

\section{Related Works}
\label{sec:related}
We begin by summarizing works that consider communication in multi-agent reinforcement learning (MARL) settings. Later we summarize works that use sim-to-real in multi-agent settings.

\textbf{Communication in MARL:}
Communication under constrained network scenarios is a well-explored area in MARL~\cite{chafii-2023-emergent,zhu_2022_comm_survey,yuan_2023_cooperative_survey,zhou_2023_survey_challenges,gronauer_2022_survey3,zhang_2021_survey4}. However many works assume perfect communication where agents can communicate and receive fresh messages from all other agents without drops or delays~\cite{d7_foerster_2016_rial_dial,c10_kim_schednet_iclr,c3_zhang_2019_VBC,zhang_2020_tmc,c6_ding_2020_i2c,c1_mao_2020_acml,c5_wang_2020_imac,c9_das_2019_tarmac,threshold_based_aaai}.
Recent studies have explored the concept of delay-aware communication, focusing on learning policies that help agents optimize their messaging by considering constraints such as limited bandwidth and noisy channels~\cite{dacom,hu_2020_event,event_triggered_PMLR,mason_2023_multi_control_comm,hu_2023_LAUREL}. However, communication between agents is assumed to be instantaneous. Any packet/observation delays are always known and predefined by a fixed communication slot rather than reflecting realistic network conditions. Such a setup also overlooks the potential impact of other agents sharing the network, which can lead to increased delays and frequent packet drops. Given such assumptions, it is never clear as to why an agent wouldn't continuously choose to send messages if there is no communication overhead or penalty. 

\textbf{Sim-to-Real in Multi-Agent Settings}:
While effective communication strategies in multi-agent settings have shown significant potential, most studies remain limited to testing within simulators or to specific task scenarios. Deploying policies trained in simulation to real-world settings poses challenges due to the gap between simulation and reality. As a result, works on transfer learning in multi-agent communication over networks is both limited and particularly challenging~\cite{gielis-2022-springer_book,prorok-k2021-arxiv,yu_ACE_asynchronous}. \textit{Domain randomization}~\cite{zero_shot_randomisation_2, sim2real_dynamic_randomization, sim2real_object_avoidance,siekmann-2020-learning,zero_shot_flow,domain_randomization_bounds} is a technique used to bridge the reality gap, further helping in the generalization of policy transfer without retraining on real-world scenarios. It involves generating many simulated environments with varied parameters and randomizing physical dynamics~\cite{sim2real_object_avoidance,siekmann-2020-learning,zero_shot_flow,zero_shot_randomisation_2} to train RL policies that can perform well when executed on real hardware like robots. Recent works have demonstrated sim-to-real transfer in multi-agent scenarios, including applications like maneuver planning in autonomous vehicles~\cite{candela_2022_IROS,yuan_202_sim2real4} and collision-free trajectory planning for UAVs~\cite{kondo_2023_robust,chen_2023_UAV2,shi_2022_simreal3}. However, domain randomization also comes with challenges. Challenges include choosing the right set of parameters to randomize~\cite{domain_random_challenge}, as otherwise, learned models may fail to generalize to real-world scenarios.  
We adopt domain randomization and propose a network simulation model, which is a low-fidelity
approximation of networks in the real world. It
abstracts away the detailed workings of wireless
networks but generates a range of communication delays, which these communicating agents might experience when deployed and executed in a real network. 

\section{Problem Setting and Formulation}
\label{sec:probSetting}
We have one or more edge-devices (agents) indexed $i=1,2,\dots$. An agent $i$ tracks a certain source process $X_i(t)$. The measurements of the process are not directly available to the agent. The measurements of processes $X_i(t)$ for every $i$ are only available at the edge-cloud. Agents can obtain measurements of their source process by querying the cloud. 

\begin{figure}
    \centering
    \includegraphics[height=0.4\linewidth]{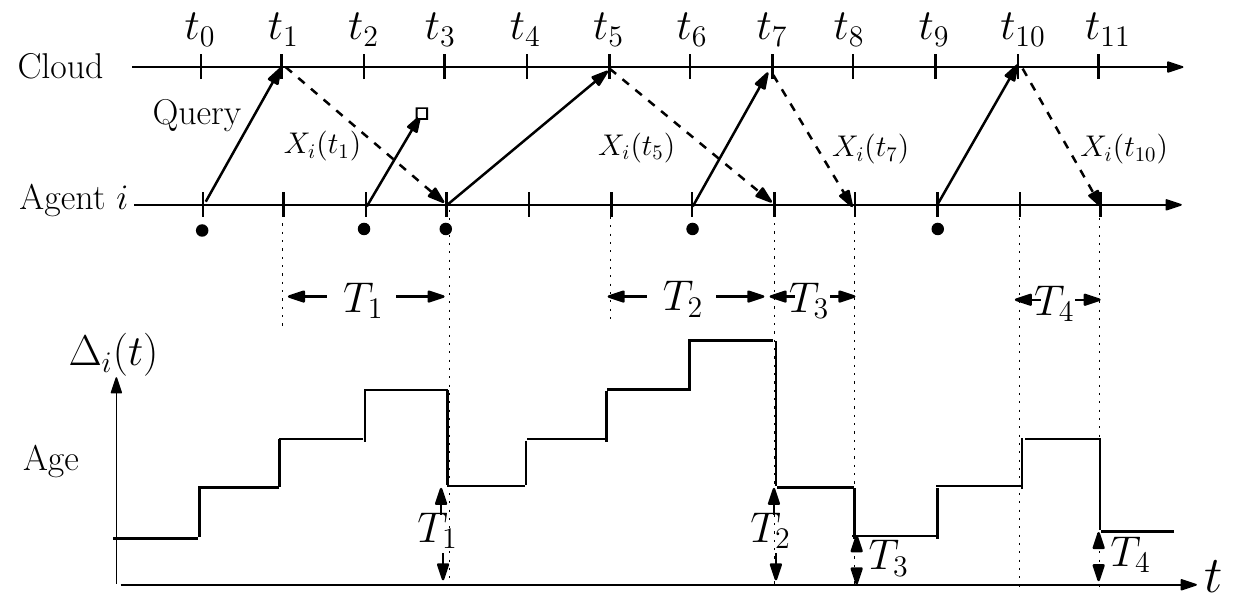}
    \caption{\small At the top is a sequence diagram of communication between agent $i$ and the edge-cloud. The agent sends a query packet to the cloud at decision instants (marked by a $\bullet$ on the agent's axis) $t_0, t_2, t_3, t_6, t_9$. The packets sent at $t_0, t_3, t_6, t_9$ are received at, respectively, $t_1, t_5, t_7, t_{10}$. The corresponding delays suffered by the query packets are $1,2,1,1$ slots. The packet sent at $t_2$ is lost and not received. For a query received at $t_k$, the cloud responds with a packet containing the most recent measurement it has from the set $\{X_i(\tau), \tau \le t_k\}$. For illustration, we assume this measurement is $X_i(t_k)$, which has the generation time $t_k$. We have $X_i(t_1), X_i(t_5), X_i(t_7), X_i(t_{10})$ received at delays of $T_1, T_2, T_3, T_4$ slots, respectively. Below in the sequence diagram, we show the age $\Delta_i(t)$ at the agent. Age jumps by a slot at every decision instant in which no packet is received by the agent. When a packet is received, age is reset to the current age of the source's measurement in the received packet. This age is the delay between its generation time and it being received by the agent. For example, $X_i(t_5)$ is received $T_3$ slots post being generated. On being received, the age at the agent is reset to $T_3$.}
    \label{fig:sequenceDiaAndAge}
    \Description{}
\end{figure}

An agent $i$ queries the cloud by sending a query data packet over a wireless network that is shared by all the agents and the edge-cloud to communicate with each other. An agent's query may or may not be received by the cloud. To exemplify, the data packet may be decoded in error by the cloud because it is received at a low signal-to-noise ratio. The data packet may be lost because of contention between the agents and the cloud communicating over the network. In case an agent's query is received by the cloud, a random amount of time (one-way delay), governed by wireless network conditions, elapses between the sending of the query packet and its reception.

Once the cloud receives a query from an agent, the query is either queued for processing behind queries that are yet to be processed and were received earlier from other agents, or in case there are no pending queries, the query from the agent is processed immediately. 
Processing of a query from agent $i$ results in the cloud sending back to $i$ the most recent measurement of the process $X_i(t)$ available at the cloud, at time $t$ of processing. The measurement is sent over the wireless network as a data packet addressed to the agent. As with any packet sent over the network, it may not be received by the agent and, if received, it suffers from a random amount of delay in the network. Figure~\ref{fig:sequenceDiaAndAge} provides an example sequence of communication between an agent and the cloud.

Given the above setting, one may want an agent to query at a high rate. A high rate of querying would ideally result in an equally high rate of measurements received from the cloud at low delays. However, given a shared wireless network, a high rate of querying by the agents could result in significant wireless contention, resulting in the cloud receiving a small fraction of the queries, often at exceedingly large delays. The challenge for every agent $i$ is to query the cloud in a manner such that, given the shared wireless network and its use by others, tracking of its source $X_i(t)$ is optimized.


We assume decision making in discrete-time. At every time instant, an agent must independently decide whether to query the cloud or not. In our setting, the agents are not cognizant of each other's presence. However, an agent is impacted by the other agents sending queries and receiving measurements over the shared wireless network. The decision to query or not by agent $i$ must not only be a function of measurements of $X_i(t)$ already received by the agent but also of the state $N(t)$ of the shared wireless network. At any time instant, the state of a wireless network includes the number of nodes sharing the network, number of packets awaiting transmission at each node, and number of packets that are actively attempting transmission. Given the distributed nature of the information that constitutes the state of a network, an agent does not know the state of the network. It must base its decision on its own experience sending query packets to the cloud and receiving measurements from it.

Our multi-agent system has one or more decision making agents and the edge-cloud. The state $S_t$ of the environment includes the state of the wireless network and the processes $X_i(t)$, one for each agent. We have $S_t = \vec{N(t)\ X_1(t)\ X_2(t)\ \ldots}$. At any decision making instant $\tau$, an agent $i$ doesn't know the state $S_\tau$. Instead it has the set $\mathcal{X}_i(\tau)$ consisting of zero or more observations of $X_i(t), t < \tau$, received from the cloud. 

In addition, we assume that the agent tracks the age $\Delta_i(t)$ of the most recent observation of its source process $X_i(t)$, received by the agent from the cloud. Figure~\ref{fig:sequenceDiaAndAge} provides example evolution of $\Delta_i(t)$, given communication between an agent and the cloud. Specifically, $\Delta_i(t) = t - \mathcal{T}_i(t)$, where $\mathcal{T}_i(t)$ is the generation time of the measurement most recently generated, among measurements in the set $\mathcal{X}_i(\tau)$ . Thus $\Delta_i(t)$ captures the staleness of the most recently generated measurement of $X_i(t)$ available at agent $i$. At time instants the agent doesn't receive a packet from the cloud or receives a packet with a measurement of $X_i(t)$ with a generation time that is older than of a measurement available at the agent, $\Delta_i(t)$ increments by $1$ slot. Otherwise, $\Delta_i(t)$ resets to the age of the measurement received by the agent. The age of the measurement is the time interval that elapses between the generation time of the measurement sent by the cloud and the agent receiving the measurement. In Figure~\ref{fig:sequenceDiaAndAge}, the received measurements have ages $T_1, T_2, T_3, \text{and } T_4$. Note that since age increases in the absence of a fresh measurement received from the cloud, infrequent querying will result in large age. Age may also be large, despite querying, in case querying results in large delays in query packets reaching the cloud or packets from the cloud reaching the agent. 

An agent $i$ would like to learn a policy $\pi_i(\mathcal{X}_i(\tau), \Delta_i(\tau))$ that at any decision instant $\tau$ returns either $0$ (do not query) or $1$ (query), as a function of the set of known observations $\mathcal{X}_i(\tau)$ and the age of the most recent observation $\Delta_i(\tau)$. Agent $i$ would like to minimize the expectation of the sum discounted tracking (estimation) squared-error
\begin{align}
    E\left[\sum_{\tau=0} \gamma^\tau (\widehat{X}_i(\tau) - X_i(\tau))^2\right],\nonumber
\end{align}
where $\widehat{X}_i(\tau)$ is the agent's estimate of $X_i(\tau)$ at time $\tau$ and $0 < \gamma < 1$ is a discounting factor. The agent calculates the estimate $\widehat{X}_i(\tau)$ as a function of the observations in the set $\mathcal{X}_i(\tau)$ and their ages at $\tau$.

\section{Network and Edge-Cloud Simulation}
\label{sec:networkAndCloudSimulationModel}
\begin{figure}[t]
    \centering
    \includegraphics[height = 0.3\linewidth]{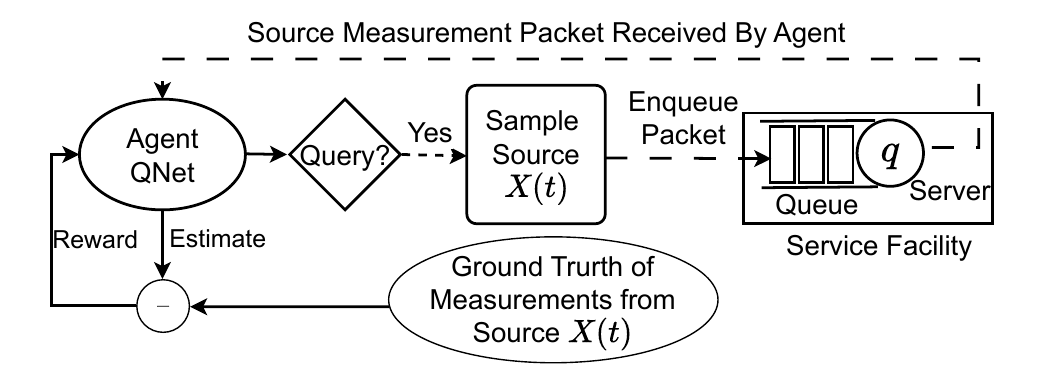}
    \caption{\small A low-cost low-fidelity simulation model to generate experiences for the agent to train its policy. The wireless network and the edge-cloud are modeled as a first-come-first-served queue with a single server. When an agent chooses to query, the source is sampled and the measurement is enqueued as a packet to be processed by queue's server. Packets are processed in the order of their arrival into the queue. A packet that enters the server takes one or more slots to finish processing, where the number of slots is a geometric random variable with parameter $0 < q \le 1$. Here $q$ is the probability of the server finishing the processing at any time instant independently of other instants. When training QNet, ground truth of source measurements is used to calculate rewards received by the agent.}
    \label{fig:neworkSim}
    \Description{}
\end{figure}

\begin{wrapfigure}{L}{0.43\linewidth}
    \centering
    \includegraphics[width=\linewidth]{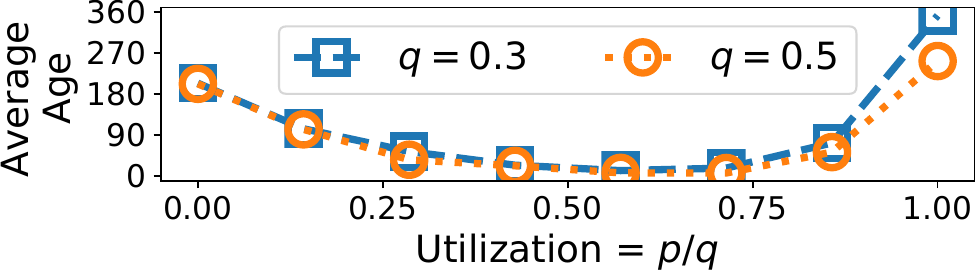}
    \caption{\small  Average age as a function of utilization $p/q$, where $0 < p \le 1$ is the packet arrival rate (average number of packet arrivals per time slot) into the service facility, shown for $q=0.3, 0.5$. Observe that the average age is large for both a very low and very high utilization.}
    \label{fig:averageAgevsp}
    \Description{}
\end{wrapfigure}

We propose a low-cost low-fidelity simulation model with a single parameter $q$. Figure~\ref{fig:neworkSim} shows an illustration of how an agent learns a policy using the simulation model. We model the network and the edge-cloud together as a service facility consisting of a single queue and a single server. The service facility processes packets in a first-come-first-served manner. 

Every decision instant the agent chooses to query the edge-cloud, a source being tracked by the agent is sampled. The resulting measurement packet enters the facility at the decision instant. If the queue is empty on the packet's arrival into the facility, the packet enters the server and begins service immediately. If a packet enters the facility when the server is busy processing another packet, the new packet waits in the queue behind other packets that arrived earlier and are yet to be serviced. The total time spent by a packet in the service facility is the sum of the waiting time and the service time, where the former is the time spent waiting in queue and the latter is the time spent in the server obtaining service.

Any packet that is in service at a decision instant finishes service in a time slot (that is by the next decision instant) with probability $q$, independently of the number of time slots it has spent in service. A packet spends at least one-time slot in service. Therefore, the service time of any packet is a geometric$(q)$ random number of time slots. On an average a packet spends $1/q$ slots in service. Note that the arrival of packets into the queue, and the resulting distribution of inter-arrival times, is entirely determined by the policy being used by the agent and is not known. 

For a fixed rate of arrival of packets into the service facility, reducing $q$ increases the number of slots servicing a packet takes on an average. As a result, new arrivals will on an average see a larger number of packets waiting in queue. They will thus spend a longer time in the service facility. Given the larger number of packets waiting in queue and a larger service time per packet, the time a packet will wait before entering the server will be larger. This followed by a larger time spent by the packet in the server. On the other hand, increasing $q$ reduces the average number of slots a packet spends getting service. As a result a new arrival spends a smaller time in the service facility. Varying $0 < q < 1$ over a large range of values enables us to simulate varied levels of network congestion, which in a real-world network could result from changing demand from users for the network resource.

For a fixed $q$, a small enough rate of arrival of packets will result in packets spend very little time waiting in queue. However, packets will be received by the agent at a small rate, that is less frequently. Infrequently received packets results in a large age of measurements at the agent. As the rate of arrival is further increased, the frequency of reception of packets by the agent will increase with each packet still suffering small enough delays. The improved frequency of reception reduces age at the agent. However, as the rate further increases, while the agent will receive packets at a higher rate, each packet would have suffered large delays due to excess queueing. This results in a large age at the agent. Note that a fixed $q$ corresponds to a fixed statistical description of the service facility and models a certain network and edge-cloud performance. An agent, when learning a policy, is able to observe how choosing to query or not changes the delays suffered by received packets and the resulting age at the agent.

Figure~\ref{fig:averageAgevsp} shows how average age at the agent changes as the probability $p$ with which a new packet arrives into the service facility increases, for a fixed $q$.

The probability $q$ models the probability of a packet being successfully transmitted in an attempt. In real networks, a packet transmission attempt may be unsuccessful because of interference from transmissions by other users or because the wireless link between the agent and cloud has a poor signal-to-noise ratio.
\section{Deep Reinforcement Learning Model}
\label{sec:drlModelQNet}
\begin{figure}[]
    \centering
    \includegraphics[width=\linewidth]{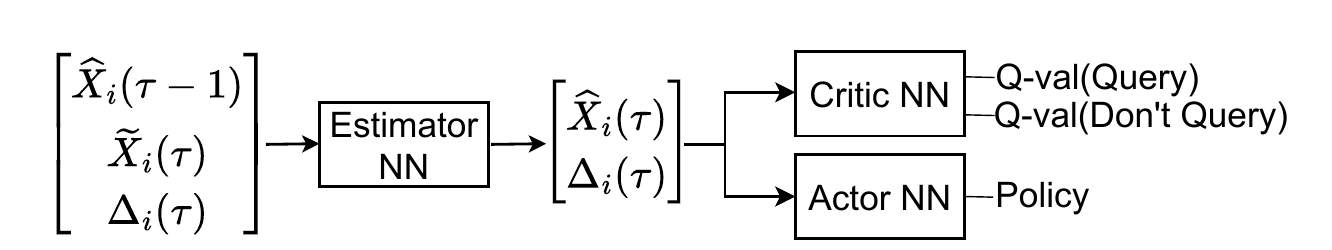}
    \caption{\small Training Pipeline of Estimator, Actor and Critic  NN.}
    \label{fig:trainingPipleine}
    \Description{}
\end{figure}
Our proposed Deep Reinforcement Learning (RL) model (\textbf{Q}uery\textbf{Net}; QNet for short) has three key neural networks, namely Estimator NN, Actor NN, and Critic NN\footnote{The Estimator NN uses an Long Short-Term Memory (LSTM) block (with $1000$ cells) followed by two fully connected layers (FC) of size 64. The Actor NN includes an input layer, two FC layers of size 256 each, followed by a softmax layer that outputs a PMF over actions. The Critic NN also has two FC layers, each of size 256.}. At decision instant $\tau$, the input to the Estimator NN is a vector containing the last calculated estimate $\widehat{X}_i(\tau - 1)$, the age $\Delta_i(\tau)$ of the most recent measurement of the source $X_i(t)$ available at the agent, and the most recent measurement $\widetilde{X}(\tau) = X_i(\tau - \Delta_i(\tau))$. At decision instant $\tau$, the Estimator NN outputs the estimate $\widehat{X}_i(\tau)$. 

Note that LSTM is a sequence model that maps a time sequence of inputs to an output sequence and has the ability to capture long-range time dependencies of the output on an input sequence. The estimate $\widehat{X}_i(\tau)$ implicitly uses the measurements received by the agent up to time $\tau$, that is the set $\mathcal{X}_i(\tau)$.

The Actor and Critic networks take as input the estimate $\widehat{X}_i(\tau)$, which is output by the Estimator NN, and the age $\Delta_i(\tau)$ at decision instant $\tau$. The Critic NN outputs the Q-values for the two actions of Query and Do not Query. The Actor NN outputs an action.

\subsection{Training the RL Model}
\begin{figure}[!ht]
	\begin{center}
		\subfloat[Query and Average Age]{\includegraphics[width = 0.4\linewidth]{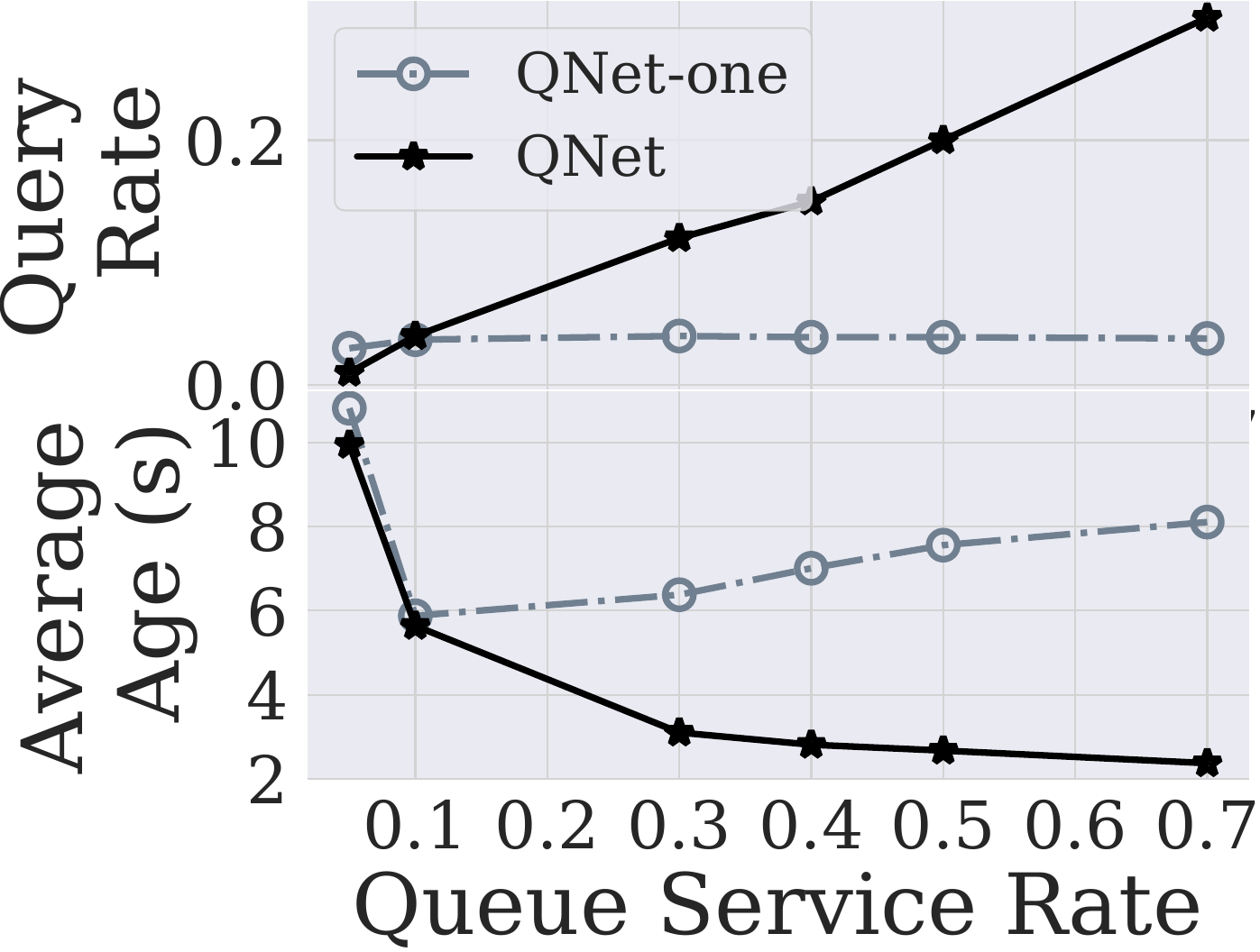}
	\label{fig:simulated_avg_age_query_rate_ver1}}
     \hspace{0.1\linewidth}
		\subfloat[Estimation Error]{\includegraphics[width = 0.4\linewidth]
{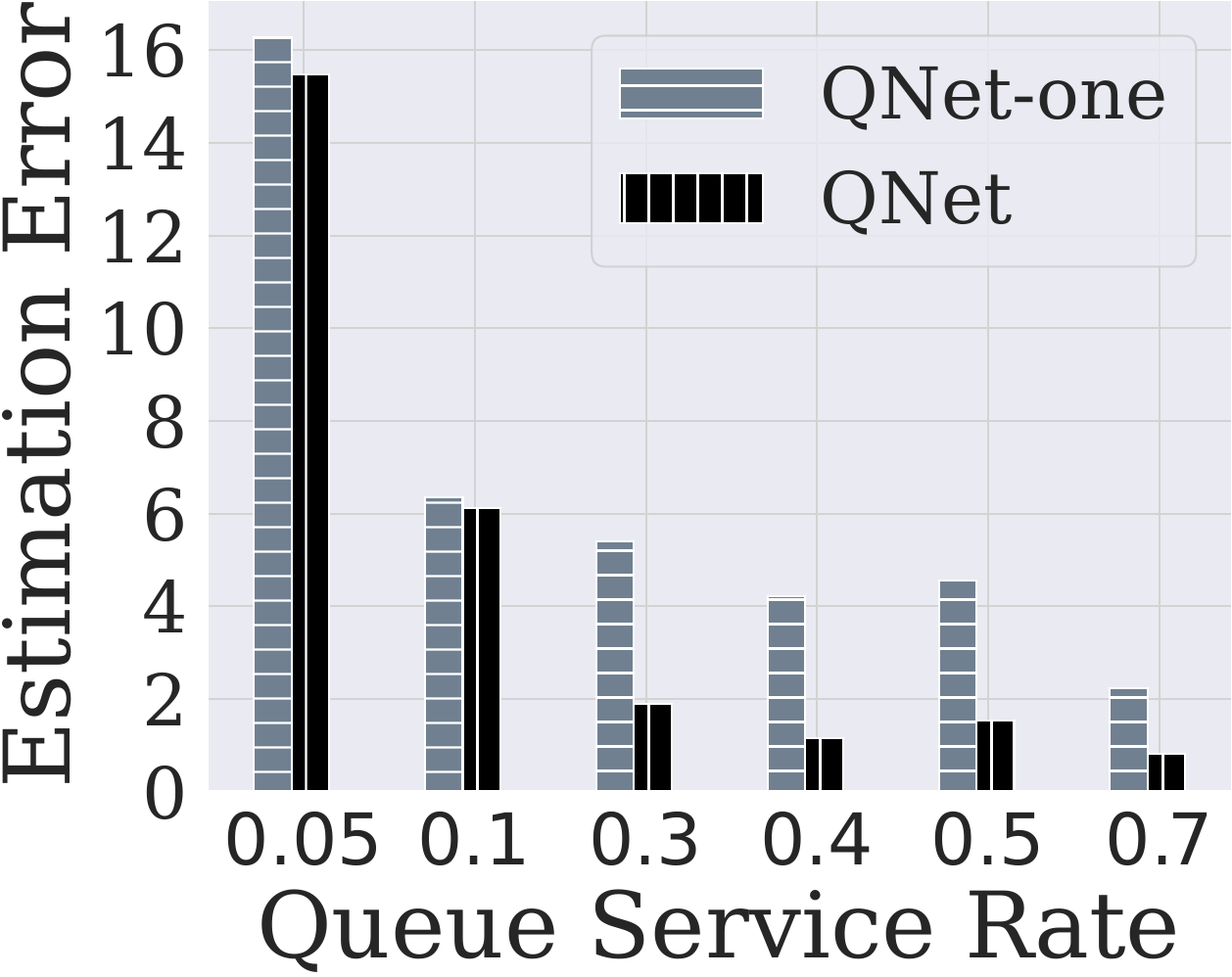}\label{fig:simulated_estimation_error_ver1}}
        \Description[]{}
		\caption{\small Comparing QNet and QNet-one. Figure (a) shows Query Rate and Average Age and (b) the estimation error, for different $q$.}
        \label{fig:threeModelsvsOne}
	\end{center}
\end{figure}
We train three RL models to cover a wide range of values of the parameter $q$ of our simulation model described in Section~\ref{sec:networkAndCloudSimulationModel}. Specifically, RL models are trained separately for episodes generated using simulations when we choose $q\in(0.05,0.1)$,  $q\in(0.1, 0.3)$, and $q\in(0.3, 1.0)$. For each subset of the values of $q$, we train the pipeline of neural nets described in Figure~\ref{fig:trainingPipleine}. We find that training different models over the subsets of values is essential to ensure that the agent chooses actions that correctly adapt to network conditions. 

Figure~\ref{fig:threeModelsvsOne} compares our chosen approach of training three models with training just one model (QNet-one) over the full range of values of $q$. The source measurements are the state of a vehicle governed by Newtonian kinematics. As is seen in Figure~\ref{fig:simulated_avg_age_query_rate_ver1}, the average rate of querying by QNet-one stays about the same over a wide range of $q$. On the other hand, QNet that trains three models, adapts the rate of querying to $q$. As one would expect, it increases the query rate with $q$. Recall that a larger $q$ corresponds to smaller average amount of time spent by a packet in service and thus a faster service rate. Not adapting the query rate results in larger average ages and a larger estimation error (see Figure~\ref{fig:simulated_estimation_error_ver1}) when using QNet-one. Larger $q$ and the resulting faster service allows QNet to query faster while keeping age low. This results in a much smaller estimation error, as $q$ increases, in comparison to the error obtained using QNet-one. As regards, QNet-one the improvement in error as $q$ increases is explained by the improved network. QNet-one, however, fails to benefit from the improved network, as is evidenced by the lack of change in query rates.

\textbf{Training Algorithm:} We use the soft actor-critic algorithm~\cite{haarnoja_2018_SAC_base} with the automatic gradient-based temperature tuning method~\cite{adaptive-dsac}. Since our action space is discrete, instead of learning a mean and covariance for an action distribution, we generate a probability mass function (PMF) over discrete actions as in~\cite{dsac}.

The agent obtains a reward after choosing an action. The reward is the negative of the error in agent's estimate of the current source measurement, suitably shifted and scaled, where the error at decision instant $\tau$ is $(\widehat{X}_i(\tau) - X_i(\tau))^2$. The reward is

     \begin{equation}
    \label{eq:reward_shift}
        \hat{r}(\tau) := r_s\left(1-\frac{(\widehat{X}_i(\tau) - X_i(\tau))^2}{\overline{r}}\right),\nonumber
    \end{equation}
    where $\overline{r}$ is the upper limit (maximum estimation error, set to $8\times 10^4$) and $r_s$ is a scaling hyperparameter, which we set to $5.0$. Note that a smaller estimation error results in a larger reward.

We use the $n$-step return, which is a well established method in literature~\cite{sutton2018reinforcement}. We set $n=60, 20, 10$, respectively, when training the RL model for $q\in(0.05,0.1)$, $q\in(0.1,0.3)$, and $q\in(0.3,1.0)$. It may take many more decision instants for a change in the agent's policy to suitably change the network condition, and the resulting rate at which packets are received by the agent from the cloud, when $q$ is small. As $q$ gets closer to $1$, and the network is able to service close to a packet every slot, a change in policy is quickly reflected in how packets are received by the agent from the cloud. The target entropy, which is a hyperparameter, is set to $0.09, 0.3, 0.6$, respectively, for the three models trained over different subsets of values of $q$. The Estimator NN used a learning rate of $10^{-4}$, while the Actor and Critic networks used a learning rate of $1.5*10^{-4}$. The replay buffer size was set to $10^6$. We used the Adam optimizer.

\textbf{Pre-Training:} When the weights for the Estimator, Actor, and Critic neural networks are initialized, the Estimator NN tends to generate highly inaccurate state estimates. This hampers the learning of policy by the Actor and Critic neural networks. We observed that this gives rise to two distinct querying behaviors: a policy biased toward frequent querying or one that does not query for extended periods of time. Such a policy completely disregards the underlying network conditions. Also it has a cascading effect in that it also worsens the estimates learnt by the Estimator NN. 

To address the above, we start with a pre-training phase during which only the weights of the Estimator NN are updated. To get around the problem of not having a policy, we query at any decision instant with probability $q/2$, where $q$ is chosen randomly for $(0,1)$ at the beginning of an episode. Note that the choice of $q/2$ ensures that the network in our simulation model is neither too lightly loaded nor very congested and allows for the agent to receive packets from the cloud at a fast enough rate and small delays, which facilitates learning of a good initial Estimator NN model.

\subsection{Training Alternate Configurations} 
\label{subsec:comparison_architectures}
We trained variants of QNet by making different choices of input for the Actor-Critic NN. 
    
    \textbf{Only Estimates, QNet-$\widehat{\mathcal{X}}$:} Here, the input consists of only the current estimate $\widehat{\mathcal{X}}(\tau)$.

    \textbf{QNet-$\lambda$:} We augment additional network statistics as in~\cite{dacom} to the input. We included the packet backlog in the queue at $\tau$, denoted by $b_i(\tau)$. This backlog serves as an indirect indicator of the network condition. The backlog $b_i(\tau)$ is updated as 
    
\begin{align} 
\label{eq:backlog}
b_i(\tau)= 
&\begin{cases}
b_i(\tau-1) + 1&  \text{if agent $i$ queries},\\\nonumber
b_i(\tau-1) - 1 & \text{if ${\mathcal{X}}(\tau)$ arrives at $i$},\\\nonumber
b_i(\tau-1) & \text{if agent $i$ does not query}.\nonumber
\end{cases}
\end{align}


In addition to the backlog, we incorporate the inter-arrival time, which is the time elapsed between consecutive arrivals of packets, containing responses to queries, at agent $i$. Therefore, the input under QNet-$\lambda$ includes the current state estimate, age of the most recently received observation, backlog, and the inter-arrival time. 


\textbf{QNet-one:} \label{one-model} We train only one model for the entire range of $q$. We set $n$-step to $50$ and target entropy to $0.2$. 




\subsubsection{Analyzing Training Performance}
\label{suppl:training_qnet}
We compare QNet with the variants described above.



\label{supl:exp_setup_supplementary}

\begin{figure}[t]
	\begin{center}
		\subfloat[Comparing QNet-One and QNet]{\includegraphics[width = 0.48\linewidth]{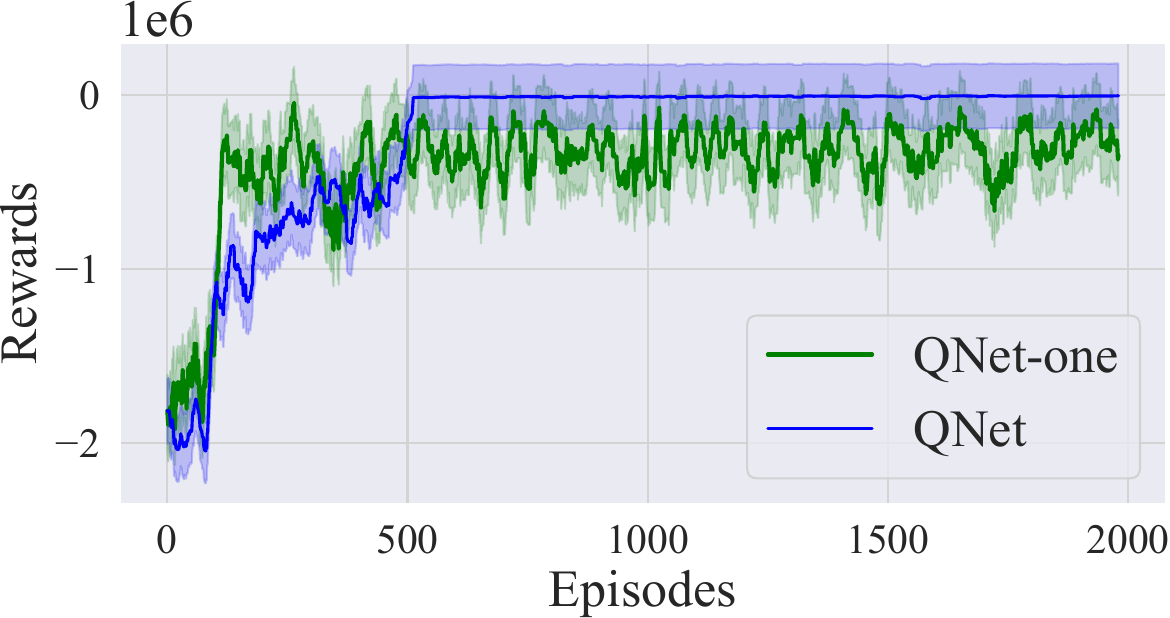}
	\label{fig:onemodelvssegmentedbuffer}}
		\subfloat[Learning curves for inputs to the QNet.]{\includegraphics[width = 0.48\linewidth]
{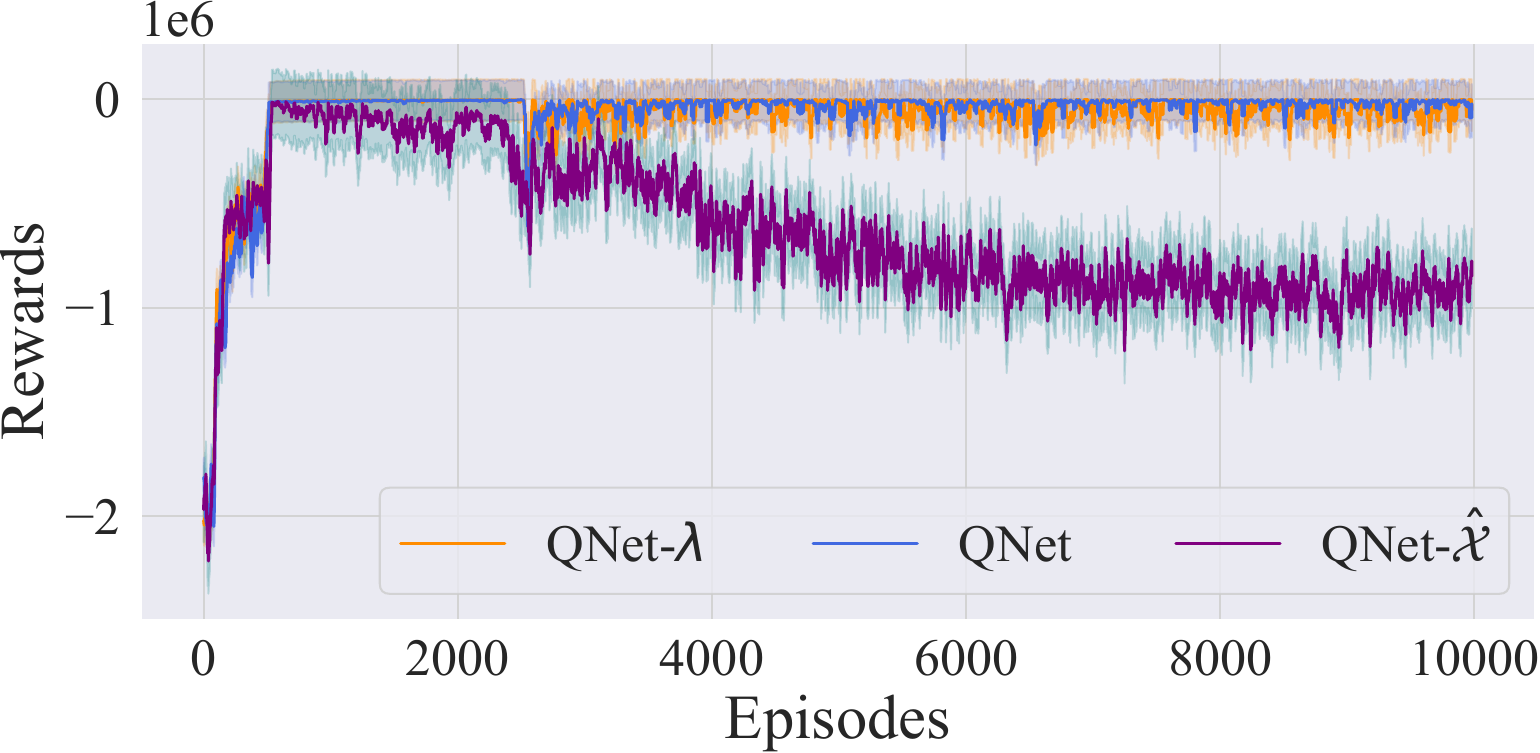}
\label{fig:comparing_inputs}}
        \Description{}
		\caption{\small Comparing training performances of QNet with QNet-One and other neural architectures with varying input configurations.}
	\end{center}
\end{figure}

\textbf{QNet-one vs. QNet:} Figure~\ref{fig:onemodelvssegmentedbuffer} shows the average of instantaneous reward over the training episodes for both QNet-one and QNet. Although QNet-one shows a steep increase in reward in the initial few episodes ($<500$), QNet achieves a better reward and is stable over long episodes.

\textbf{Variants of QNet:}
We compare QNet, QNet-$\lambda$ and QNet-$\widehat{\mathcal{X}}$ and investigate whether incorporating additional network information contributes to faster training and convergence. Figure~\ref{fig:comparing_inputs} demonstrates the instantaneous reward against the training episodes for QNet and its variants. We observe that the rewards are close for the QNet and QNet-$\lambda$, with a slight improvement for the QNet model. On the contrary, it is evident that the QNet-$\widehat{\mathcal{X}}$ does not perform well, as it has a declining reward over the training episodes. Therefore, incorporating the age of the most recently received observation proves to be beneficial in facilitating the learning process and improving the overall performance of the query agent. In conclusion, our proposed model QNet does at least as well or better than its variants.

\subsection{RL model for deployment in an unknown network:} 
\begin{figure}[t]
    \centering
    \includegraphics[width=1.0\linewidth]{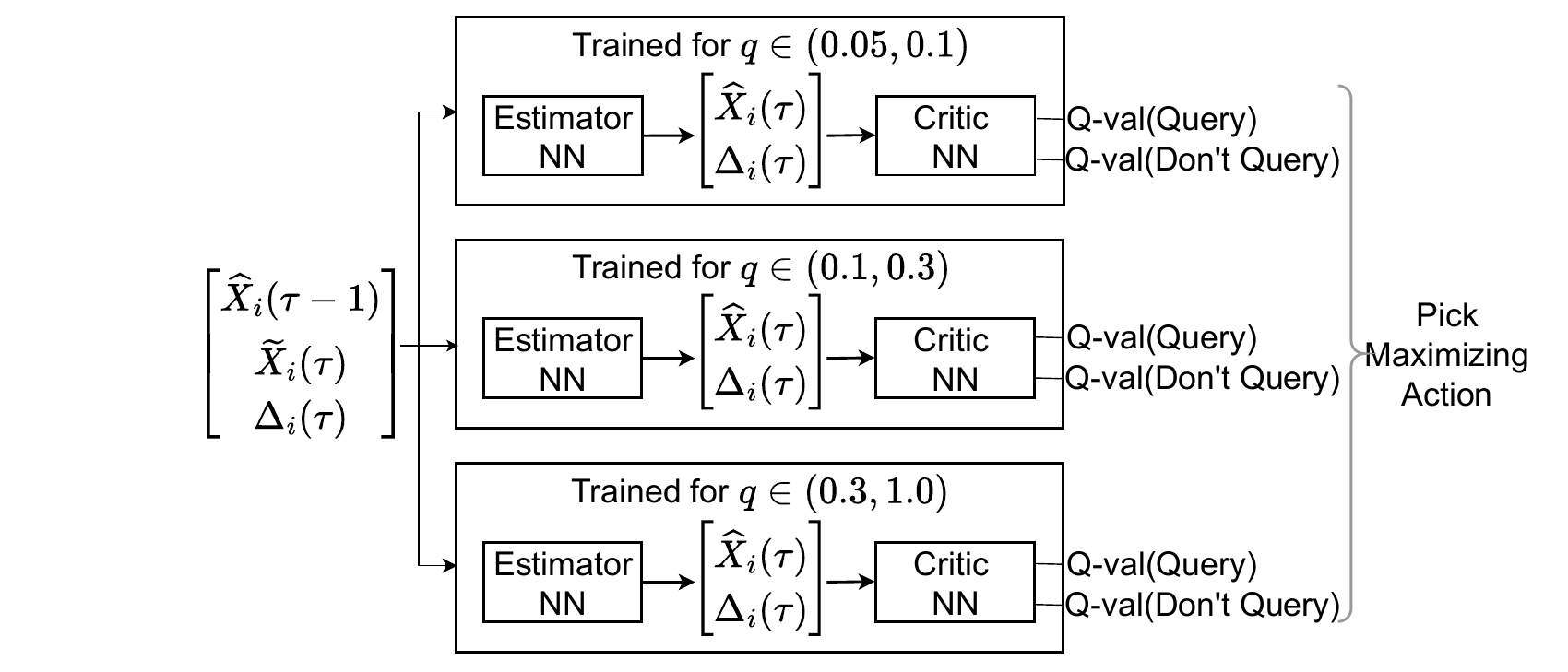}
    \caption{\small Selecting actions during execution in an unknown network environment using three neural networks.}
    \label{fig:testingPipleine}
    \Description{}
\end{figure}
Figure~\ref{fig:testingPipleine} shows how we use the three models, post training, to select actions in an unknown network. We use the trained Estimator and Critic NN from each of the three models to output Q-values for the actions Query and Do not Query. Given the set of six Q-values, we pick the action corresponding to the maximum Q-value in the set.

\section{Transfer to Real Wireless Networks}
\label{sec:experiments}
We evaluate the model QNet, which we trained using the simulation model (Figure~\ref{fig:neworkSim}), on a real wireless ($802.11a$, WiFi) network.

\subsection{Real WiFi Networks Using the ORBIT Grid} 
\label{sec:orbitGridDesc}
We used the ORBIT\footnote{https://www.orbit-lab.org/} wireless network grid to experiment with real WiFi networks. ORBIT houses a large number of WiFi radios deployed as a grid with adjacent nodes along columns and rows $1$ m apart, over an area of about $20$m $\times 20$m. 
The WiFi radios (Atheros chipsets) and supported networking protocols are real-world and commonly found in off-the-shelf home/office WiFi devices and Internet-of-Things (IoT) devices. The key benefit is that ORBIT allows an experimenter to remotely configure the wireless radios in the grid as per the requirements of an experiment. The dense deployment of radios in the ORBIT grid makes it easy to emulate wireless network contention (up to packet error rates as high as $60\%$, in our experiments) and related network congestion (one-way link delays as small as $10$ ms to as large as $2$ seconds) via appropriately designed experiments. Contention and congestion are key challenges that real-world networks face, given the scarcity of wireless spectrum and an increasing demand for the same. The physical spread of the ORBIT grid has received signal strengths vary from a high of $-40$ dBm to smaller than $-70$ dBm. With regards to an application using a wireless network, signal strength, wireless link packet error rates and delays, are key indicators of wireless network performance. ORBIT allows us to experiment with a wide range of these indicators with real-world radios and networking.


\subsection{WiFi Network, Agents, Sources, Edge-Cloud}
\label{subsec:setup_orbit}
\begin{figure*}[h]
\begin{center}
\includegraphics[width = \linewidth] {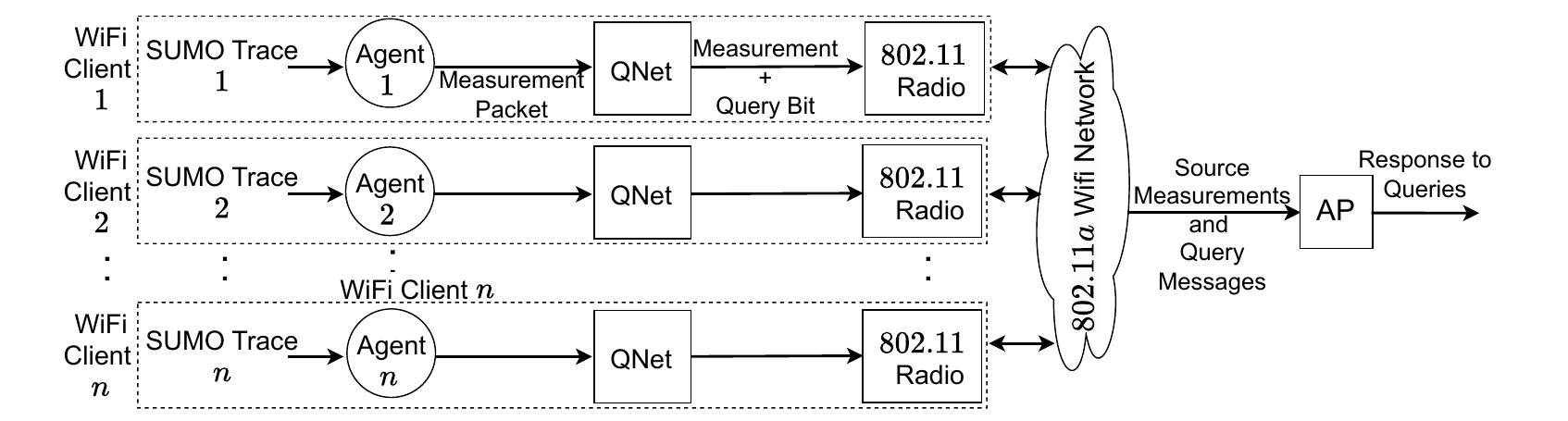}
\Description[ORBIT Setup]{}
\caption{\small {ORBIT Setup: Source $i$ uses SUMO trace $i$ to generate measurement packets and set the query bit to $0$ or $1$ in the packet. It uses WiFi client $i$ to send packets over the WiFi network to the AP. The AP then sends responses to queries. In our experiments, we vary the number of sources from $5$ to $50$.}}
\label{fig:orbit_setup}
\end{center}
\end{figure*}
For all our experiments we configured a WiFi network in the infrastructure mode, where we had all agents connect as WiFi clients to a single WiFi access point (AP). In the infrastructure mode all communication happens via the AP. A certain node on the ORBIT grid was configured as the AP using the \texttt{hostapd} utility. Every agent was a distinct ORBIT node with its own WiFi radio. The AP and the agents used the $802.11$a WiFi standard to communicate with each other. We configured them to use a fixed channel in the $5$ GHz band. We set the physical layer rate to $6$ Mbps. The AP and the agents were time synchronized using the Network Time Protocol (NTP). The AP also served as the edge-cloud. Each agent executed the trained QNet RL model every decision instant. Note that the decision instants are not synchronized across the agents. Such synchronization is not required and is infeasible in practice.

In our simulation model, we assumed the presence of a source readily available to the edge-cloud. In practice, however, the edge-cloud must receive data from edge-devices like sensors and this communication also happens using a wireless network. In our experiments, we had each agent ORBIT node also act as a sensor sending messages to the edge-cloud (AP), using the same wireless network. The sensor messages were generated by each agent ORBIT node once every $0.1$ second and included a timestamp that recorded the time the message was generated. In addition, every agent made a decision to query or not every $0.1$ second. The choice of action made by an agent was sent as part of the sensor message, generated at the decision instant, to the edge-cloud. If the choice was query, it indicated to the edge-cloud that the agent wanted to be sent a measurement of its source of interest. The edge-cloud maintained a separate queue for sensor messages received from each agent. The sensor messages constituted the sources at the edge-cloud, one source per agent. The edge-cloud processed queries from agents in a first-come-first-served manner. Given a query from an agent for a measurement from a particular source, the edge-cloud sent a packet to the agent with the most recent measurement it had from the source together with its generation timestamp. Recall the generation timestamp enables calculation of age of the measurement.


Every sensor, one at every agent ORBIT node, generated messages that contained the state of a certain vehicle moving in a traffic simulation created using the micro traffic simulator SUMO (Simulation of Urban MObility)~\cite{sumo}. Our traffic simulation had a square road network comprising of a single junction at the center of the square, with each road having $6$ lanes and up to $300$ vehicles. The road length was kept constant at about $5$ km with an intersection radius of $25$ m, and the maximum velocity of any vehicle was maintained at $10$ m/s. The driving behavior of any vehicle in the simulator was governed using the Intelligent Driver Model (IDM)~\cite{idm}. The $LC2013$ lane change controller was used to allow vehicles to change lanes. Traffic lights at the intersections further helped introduce perturbations in the traffic. All other traffic simulation parameters were set to as in~\cite[Section~$3$]{zero_shot_flow}. We randomly mapped a sensor to a vehicle in the simulation. Also, we mapped every agent to a source (sensor) in a manner that every agent is mapped to a distinct source and the source corresponds to a sensor not on the agent's own ORBIT node. A vehicle's state includes its position and velocity along the $x$ and $y$ coordinates. The SUMO generated vehicle states also served as a ground truth that helped evaluate the performance of QNet and other chosen baseline policies over the configured WiFi network. 

We performed experiments with the number of agents set to $5$, $25$, $30$, $40$, and $50$. We set the measurement packet size to $1024$ bytes\footnote{We don't restrict ourselves to the actual size of a data payload containing position, velocity, and acceleration information. The measurements in practice come with signature information for authentication purposes, resulting in larger payloads of about $200$ bytes in size. In general, the measurements could be images or LIDAR pointclouds, resulting in even larger data payloads.}. This allows us to evaluate QNet over a wide range of wireless network conditions. Figure~\ref{fig:wirelessNetworkCharacteristics} summarizes the RSS (received signal strength), the packet error rate, the round-trip times, and the retry rates, we observed during our experiments. A higher RSS typically corresponds to a high rate of successfully decoding received packets. The AP is the blue square. The locations of the other colored squares are where the agents ($50$ in the figure) were on the ORBIT grid. The color of the squares indicates the RSS. We see a range of more than $20$ dB. The retry percentage is the average fraction of packets that were retransmitted (repeat transmissions by the WiFi protocol because of earlier failed attempts). Varying the number of agents has us observe close to no retires to a significant fraction of retries. Retries can occur when contention for the wireless channel is high or because of a low RSS. Correspondingly, we see a wide range of round-trip-times (RTT) and packet error rates (PER). The PER captures the fraction of queries from any agent for which the querying agent didn't receive any response from the edge-cloud. A response may not be received either because the query packet from the agent is lost or because the measurement packet sent by the edge-cloud is lost. The RTT is the time that elapses between the agent sending a query and receiving a response to it from the edge-cloud.

\subsection{Efficacy of the Sim-To-Real Transfer}
\begin{figure}[t] 
	\begin{center}
		\includegraphics[height =0.25\linewidth]{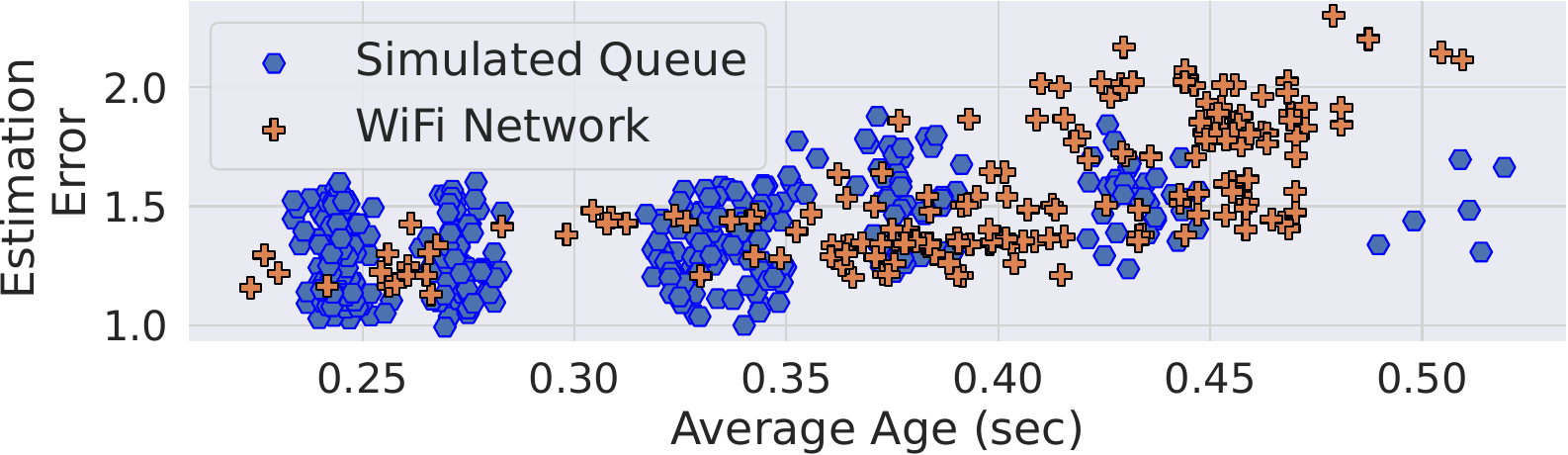}
		\caption{\small A scatter plot of average age and average estimation error for experiments using real WiFi networks overlaid with a scatter obtained using the network simulation model.}
		\label{fig:sim2real}
	\end{center}
 \Description{}
\end{figure}

\begin{table}[]
\fontsize{9pt}{9pt}\selectfont
\centering
\caption{\small Mean and standard deviation of estimation error when using QNet, shown for different average age bins.}
\label{tab:sim2real_performance}
\begin{tabular}{@{}cccccc@{}}%
\toprule
\multicolumn{6}{c}{\textbf{Estimation Error (Average $\pm$ Standard Deviation)}} \\ \midrule
\textbf{\begin{tabular}[c]{@{}c@{}}Average Age\\ (sec)\end{tabular}} & \textbf{Simulated} & \multicolumn{1}{c|}{\textbf{Real WiFi}} & \multicolumn{1}{c}{\textbf{\begin{tabular}[c]{@{}c@{}}Average Age\\ (sec)\end{tabular}}} & \multicolumn{1}{c}{\textbf{Simulated}} & \multicolumn{1}{c}{\textbf{Real WiFi}}  \\ \midrule


{[}0.2, 0.25) & 1.12 $\pm$ 0.16 & \multicolumn{1}{c|} {1.20 $\pm$ 0.05} & {[}0.25, 0.3) & 1.22 $\pm$ 0.14 & \multicolumn{1}{c}  {1.27 $\pm$ 0.09}   \\

{[}0.3, 0.35) & 1.34 $\pm$ 0.12 & \multicolumn{1}{c|} {1.40 $\pm$ 0.08} & {[}0.35, 0.4) &  1.39 $\pm$ 0.12 & \multicolumn{1}{c}  {1.39 $\pm$ 0.14}   \\


{[}0.4, 0.45) & 1.48 $\pm$ 0.14 & \multicolumn{1}{c|} {1.67 $\pm$ 0.22} & {[}0.45, 0.5) &  1.53 $\pm$ 0.13 & \multicolumn{1}{c}  {1.80 $\pm$ 0.22}  \\

\bottomrule
\end{tabular}%
\end{table}

The simulation model allows us to test QNet over different network conditions by varying $q$ over $(0.05,1)$. At test time, we generate multiple episodes, where each episode is generated using a randomly chosen $q$. In our experiments using ORBIT, changing the numbers of agents results in different network conditions. 

Figure~\ref{fig:sim2real} shows the relationship between average estimation error and average age obtained from simulations and the ORBIT experiments. For simulations, the averages were calculated per episode. We ran $100$ episodes with $40000$ time slots each. For experiments using WiFi networks the averages were calculated per agent per experiment. Observe that the scatter plots overlap quite well, indicating that the QNet model trained in simulation transferred well to experiments with real WiFi networks.

Table~\ref{tab:sim2real_performance} zooms in to show the efficacy of the transfer over different bins of average age. For each bin, we tabulate the average estimation error and the standard deviation. We see the averages and the standard deviations obtained from experiments on ORBIT are almost as good as those obtained in simulations. Our experiments show that domain randomization together with the proposed simulation model trains the RL model QNet in a manner that transfers well to real WiFi networks without any addition training over the WiFi network, that is QNet transfers well in a zero-shot manner. 

\subsection{Evaluation and Comparison With Baselines}
\label{subsec:orbit_baselines_policies}

\begin{table*}[!h]
\centering
\caption{Mean and standard deviation of estimation error obtained using QNet and the baselines. For threshold based and probabilistic querying, boxes highlight their variants that resulted in the lowest error for the chosen number of agents. The \textbf{highlighted} error values in a row mark the best performing policy (smallest mean error) between QNet and Always Query for the number of agents.}
\label{tab:orbit_estimation_error}
\resizebox{\textwidth}{!}{%
\begin{tabular}{@{}c|c|cccccc|cccc|c@{}}
\toprule
\multirow{2}{*}{\# Agents} & \multirow{2}{*}{\textbf{\begin{tabular}[c]{@{}c@{}}Always \\ Query\end{tabular}}} & \multicolumn{6}{c|}{\textbf{\begin{tabular}[c]{@{}c@{}}Threshold Based Querying\end{tabular}}} & \multicolumn{4}{c|}{\textbf{\begin{tabular}[c]{@{}c@{}}Probabilistic 
 Querying ($\sigma(.)$)\end{tabular}}} & \multicolumn{1}{l}{\multirow{2}{*}
{\textbf{QNet}}} \\ \cmidrule(lr){3-12}
 &  & \textbf{$\delta=3$} & \textbf{$\delta=2$} & \textbf{$\delta=1$} & \textbf{$\delta=0.5$} & \textbf{$\delta=0.25$} & \textbf{$\delta=0.1$} & \textbf{$\sigma(e_r)$} & \textbf{$\sigma(\frac{e_r}{N})$} & \textbf{$\sigma(0.5\,\frac{e_r}{N})$} & \textbf{$\sigma(0.33\,\frac{e_r}{N})$} &  \\ \midrule
 
5 & \begin{tabular}[c]{@{}c@{}}\textbf{0.6±0.1}\end{tabular} & \begin{tabular}[c]{@{}c@{}}1.9±  0.2\end{tabular} & \begin{tabular}[c]{@{}c@{}}1.6±  0.3\end{tabular} & \begin{tabular}[c]{@{}c@{}}1.5± 0.4\end{tabular} & \begin{tabular}[c]{@{}c@{}}1.4 ±  0.5\end{tabular} & \begin{tabular}[c]{@{}c@{}}1.3±  0.5\end{tabular} & \begin{tabular}[c]{@{}c@{}}\fbox{1.2±  0.5}\end{tabular} & \begin{tabular}[c]{@{}c@{}}1.2±  0.5\end{tabular} & \begin{tabular}[c]{@{}c@{}}\fbox{1.0±  0.2}\end{tabular} & \begin{tabular}[c]{@{}c@{}}1.1±  0.3\end{tabular} & \begin{tabular}[c]{@{}c@{}}1.4±   0.3\end{tabular} & \begin{tabular}[c]{@{}c@{}}1.0± 0.1\end{tabular} \\ \midrule

25 & \begin{tabular}[c]{@{}c@{}}\textbf{1.3± 0.3}\end{tabular} & \begin{tabular}[c]{@{}c@{}}2.0 ± 0.3\end{tabular} & \begin{tabular}[c]{@{}c@{}}1.7±  0.4\end{tabular} & \begin{tabular}[c]{@{}c@{}}1.5±  0.5\end{tabular} & \begin{tabular}[c]{@{}c@{}}1.3±  0.6\end{tabular} & \begin{tabular}[c]{@{}c@{}}\fbox{1.2±  0.6}\end{tabular} & \begin{tabular}[c]{@{}c@{}}1.2±  0.6\end{tabular} & \begin{tabular}[c]{@{}c@{}}\fbox{1.3±  0.5}\end{tabular} & \begin{tabular}[c]{@{}c@{}}1.5±  0.6\end{tabular} & \begin{tabular}[c]{@{}c@{}}1.5±  0.5\end{tabular} & \begin{tabular}[c]{@{}c@{}}1.6±  0.5\end{tabular} & \begin{tabular}[c]{@{}c@{}}\textbf{1.3± 0.4}\end{tabular} \\ \midrule

30 & \begin{tabular}[c]{@{}c@{}}2.4± 0.5\end{tabular} & \begin{tabular}[c]{@{}c@{}}2.0 ± 0.3\end{tabular} & \begin{tabular}[c]{@{}c@{}}1.8± 0.4\end{tabular} & \begin{tabular}[c]{@{}c@{}}\fbox{1.6± 0.5}\end{tabular} & \begin{tabular}[c]{@{}c@{}}1.6± 0.7\end{tabular} & \begin{tabular}[c]{@{}c@{}}1.8± 0.6\end{tabular} & \begin{tabular}[c]{@{}c@{}}1.6± 0.6\end{tabular} & \begin{tabular}[c]{@{}c@{}}1.6± 0.6\end{tabular} & \begin{tabular}[c]{@{}c@{}}\fbox{1.4± 0.6}\end{tabular} & \begin{tabular}[c]{@{}c@{}}1.5± 0.5\end{tabular} & \begin{tabular}[c]{@{}c@{}}1.8± 0.5\end{tabular} & \begin{tabular}[c]{@{}c@{}}\textbf{1.5± 0.4}\end{tabular} \\ \midrule

40 & \begin{tabular}[c]{@{}c@{}}3.7± 0.5\end{tabular} & \begin{tabular}[c]{@{}c@{}}\fbox{2.1±  0.3}\end{tabular} & \begin{tabular}[c]{@{}c@{}}2.6± 0.7\end{tabular} & \begin{tabular}[c]{@{}c@{}}2.9 ±  0.5\end{tabular} & \begin{tabular}[c]{@{}c@{}}2.7± 0.5\end{tabular} & \begin{tabular}[c]{@{}c@{}}2.7±  0.5\end{tabular} & \begin{tabular}[c]{@{}c@{}}3.0±  0.5\end{tabular} & \begin{tabular}[c]{@{}c@{}}\fbox{2.6± 0.5}\end{tabular} & \begin{tabular}[c]{@{}c@{}}2.9±  0.6\end{tabular} & \begin{tabular}[c]{@{}c@{}}3.3± 0.8\end{tabular} & \begin{tabular}[c]{@{}c@{}}3.3± 0.7\end{tabular} & \begin{tabular}[c]{@{}c@{}}\textbf{2.0± 0.3}\end{tabular} \\ \midrule

50 & \begin{tabular}[c]{@{}c@{}}31.3± 9.2\end{tabular} & \begin{tabular}[c]{@{}c@{}}\fbox{29.2 ±  8.9}\end{tabular} & \begin{tabular}[c]{@{}c@{}}30.2± 13.7\end{tabular} & \begin{tabular}[c]{@{}c@{}}31.8± 15.6\end{tabular} & \begin{tabular}[c]{@{}c@{}}35.5± 11.7\end{tabular} & \begin{tabular}[c]{@{}c@{}}41.4 ± 10.7\end{tabular} & \begin{tabular}[c]{@{}c@{}}46.1± 9.1\end{tabular} & \begin{tabular}[c]{@{}c@{}}31.3±  7.4\end{tabular} & \begin{tabular}[c]{@{}c@{}}38.6±  8.3\end{tabular} & \begin{tabular}[c]{@{}c@{}}32.0±  12.2\end{tabular} & \begin{tabular}[c]{@{}c@{}}\fbox{28.8±  7.9}\end{tabular} & \begin{tabular}[c]{@{}c@{}}\textbf{25.0±  2.3}\end{tabular} \\ \bottomrule
\end{tabular}%
}
\end{table*}

\begin{figure*}[t]
\begin{center}
\subfloat[\small{RSS (dBm)}]{\includegraphics[height = 0.15\linewidth] {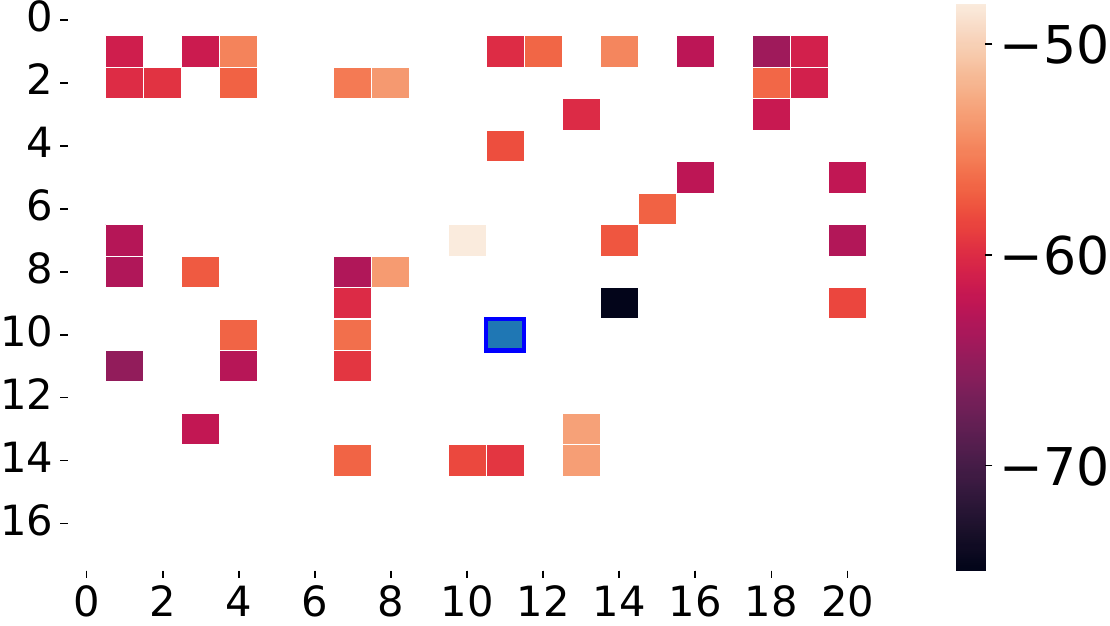}
\label{fig:rss}}
\subfloat[\small{PER ($\%$)}]{\includegraphics[height = 0.15\linewidth] {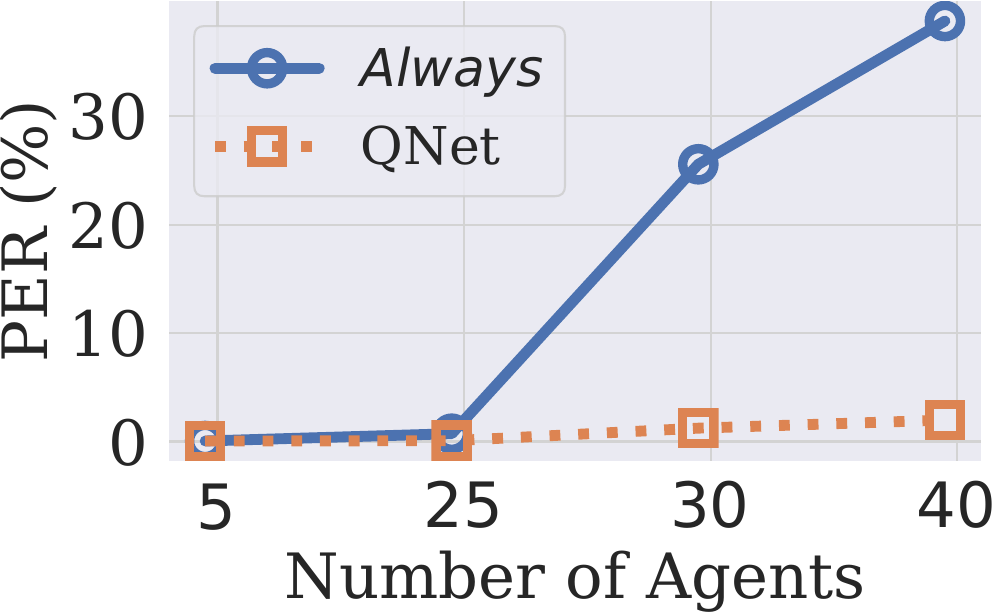}
\label{fig:per_orbit}}
\subfloat[\small{RTT (s)}]{\includegraphics[height = 0.15\linewidth] {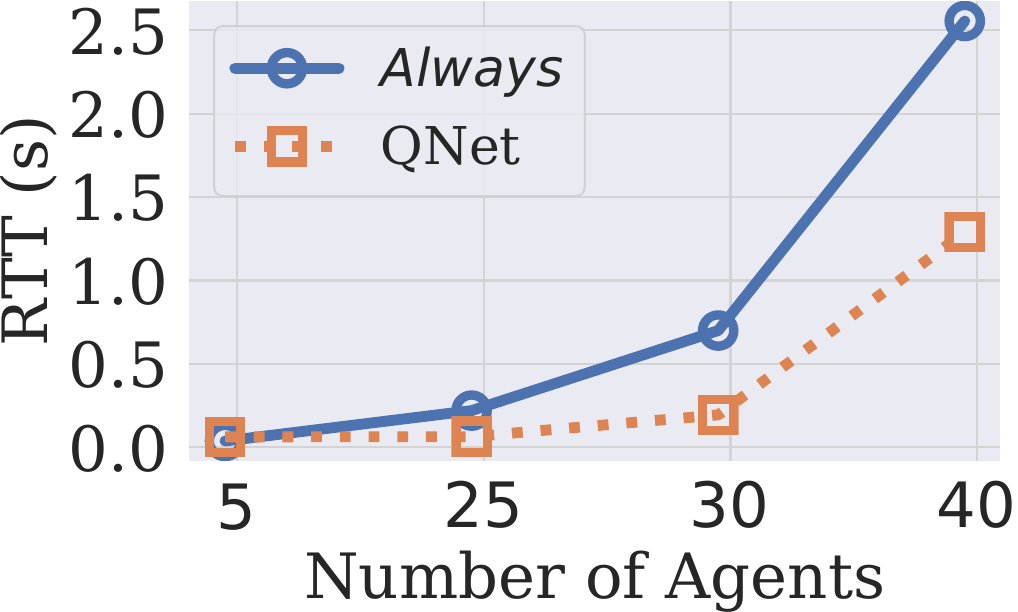}
\label{fig:rtt_orbit}}
\subfloat[\small{Retries $(\%)$}]{\includegraphics[height = 0.15\linewidth] {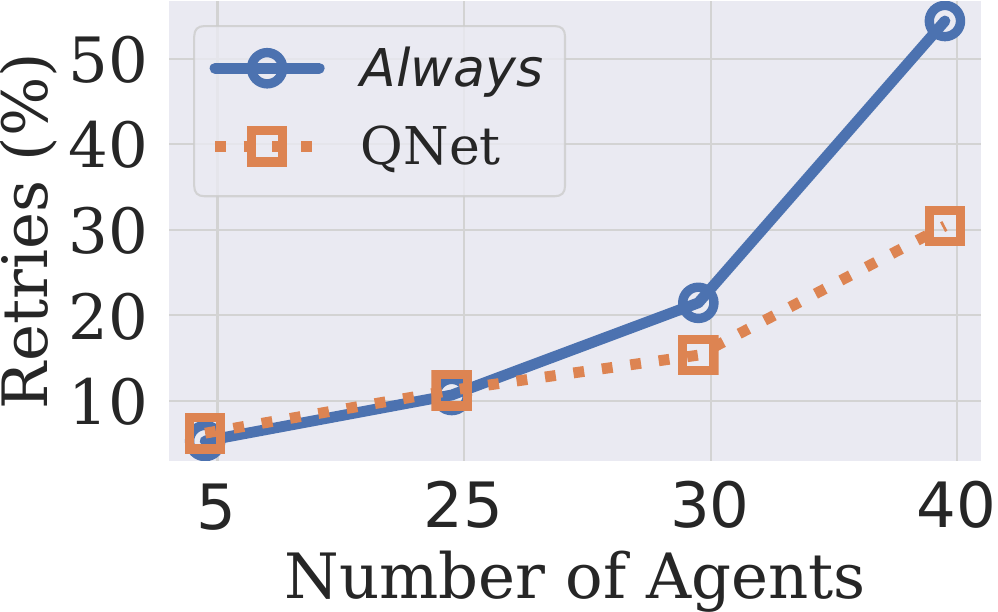}
\label{fig:retries_exp}}
\Description[ORBIT Wireless Characteristics]{}
\caption{\small (a) Received Signal Strength in dBm, (b) Packet Error Rate (\%) , (c) RTT (s), and (d) Retry Rate (\%), with (b), (c) and (d) plotted as a function of different numbers of agents. The AP is placed at the center (blue rectangle in (a)). }
\label{fig:wirelessNetworkCharacteristics}
\end{center}
\end{figure*}

We compared QNet with three policies, which we describe next.

\textbf{Always Query}: Under this policy, an agent queries at every decision instant, which is every $0.1$ seconds in our WiFi experiments. The policy is agnostic to network conditions and is trivial to implement. Such an approach to communication is seen in many works~\cite{sukhbaatar_2016_commnet,c9_das_2019_tarmac,ic3net}, and often assumes perfect network conditions, that is a message is always received and without much delay. 
    
\textbf{Threshold Based Policy}: The authors in~\cite{threshold_based_aaai} proposed model-based communication, where agents send messages only when the estimation error exceeds a heuristically chosen threshold $\delta$. We implemented a similar policy, where each agent $i$ calculated the instantaneous estimation error and queried if the error exceeded a specific threshold. We evaluated threshold values of $3, 2, 1, 0.5, 0.25, \text{and}\, 0.1$ in our experiments. The smaller the threshold, the more frequent the querying. It is worth noting that calculation of error requires the agent to have immediate access to the true state of its source. This is of course not possible in practice and if the true state were available, it would make estimating it pointless. We are able to access the true state because our state is obtained from simulated SUMO traces. Our threshold based policies, while impractical, help shed light on the performance of QNet, as we will see later. 

\textbf{Probabilistic Querying ($\sigma$ based)}: We use the sigmoid function in conjunction with the estimation error at a decision instant. The error, scaled appropriately, is an argument to the sigmoid function $\sigma(x) = 1/(1 + exp(-x))$. The output of the sigmoid serves as the probability with which an agent queries. We evaluate variants that use $\frac{e_r}{N}, \frac{1}{2}\frac{e_r}{N}, \frac{2}{3} \frac{e_r}{N}$ as input to $\sigma(.)$, where $N$ is the number of agents sharing the network and $e_r$ is the estimation error at the decision instant. The policies assume the knowledge of the total number of agents and, like the threshold based policies, of the estimation error. In essence, an agent that observes a larger error will query with higher probability. The factor $1/N$ makes the probability inversely proportional to the number of agents and allows for better sharing of the network.

\begin{figure*}[t]
\begin{center}
\subfloat[\small{Estimation Error}]{\includegraphics[width = 0.3\linewidth] {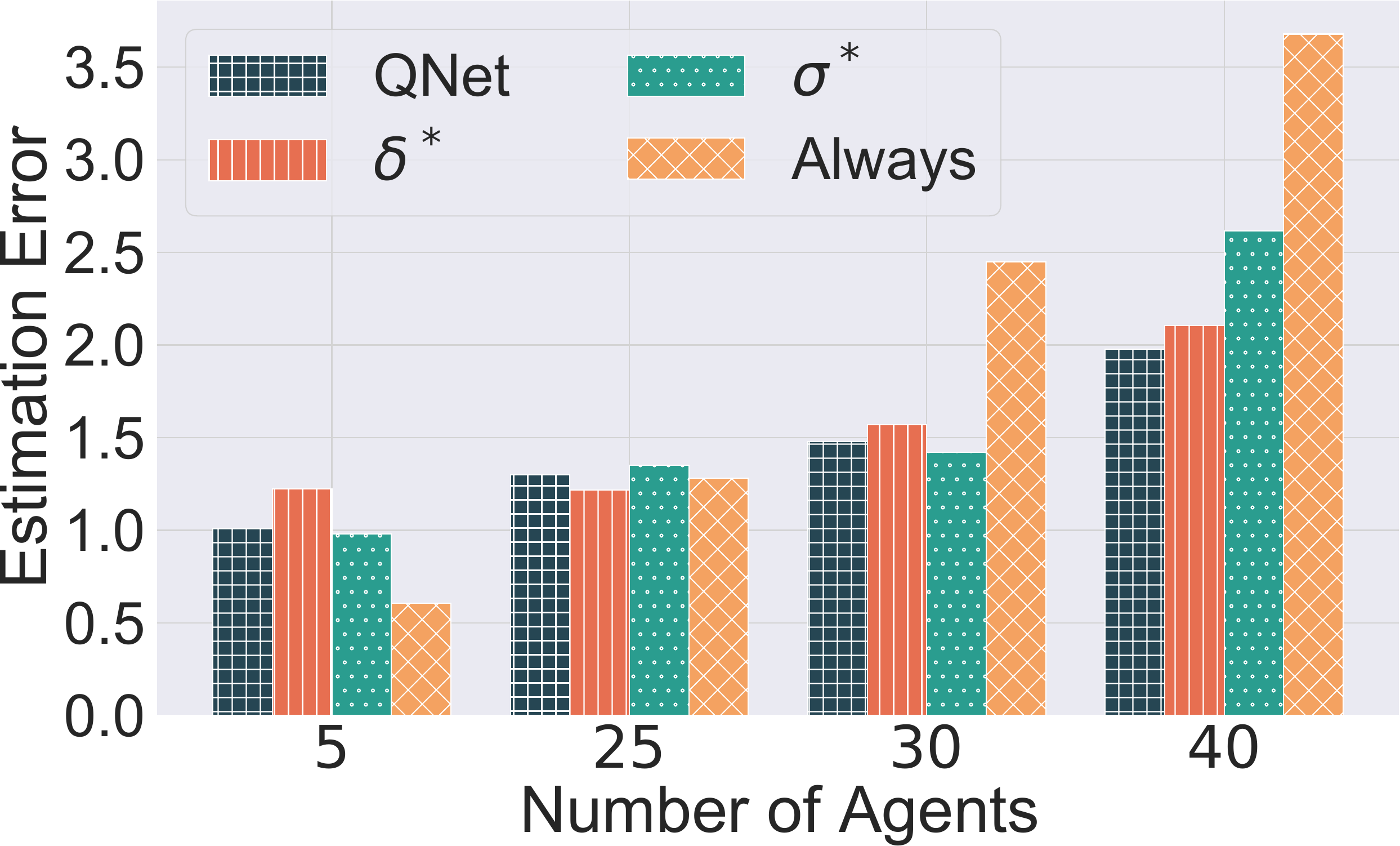}
\label{fig:bar_estimation_orbit}}
\hfill
\subfloat[\small{Query Rates (s)}]{\includegraphics[width = 0.3\linewidth] {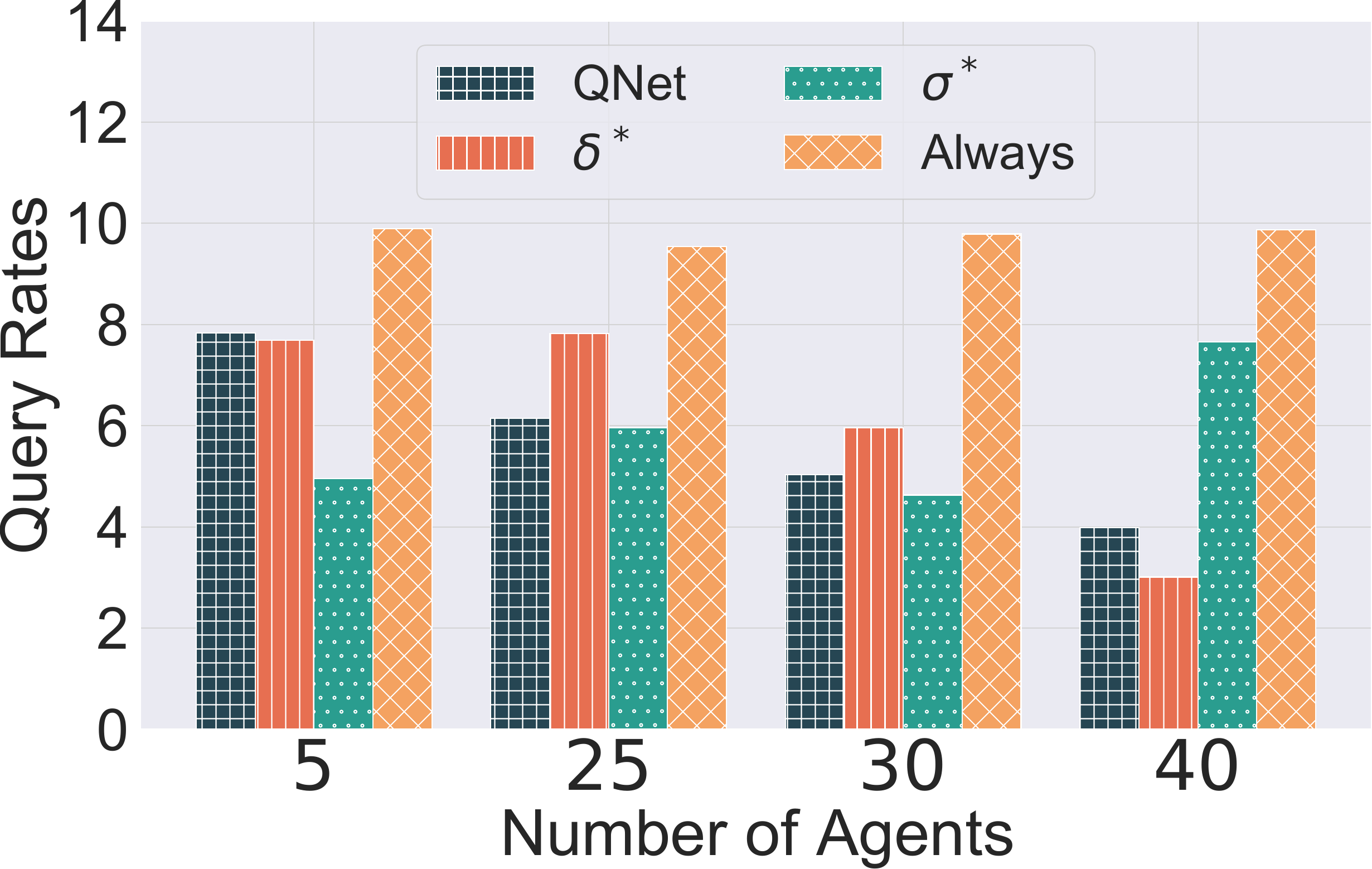}
\label{fig:bar_age_query}}
\hfill
\subfloat[\small{Average Age (s)}]{\includegraphics[width = 0.3\linewidth] {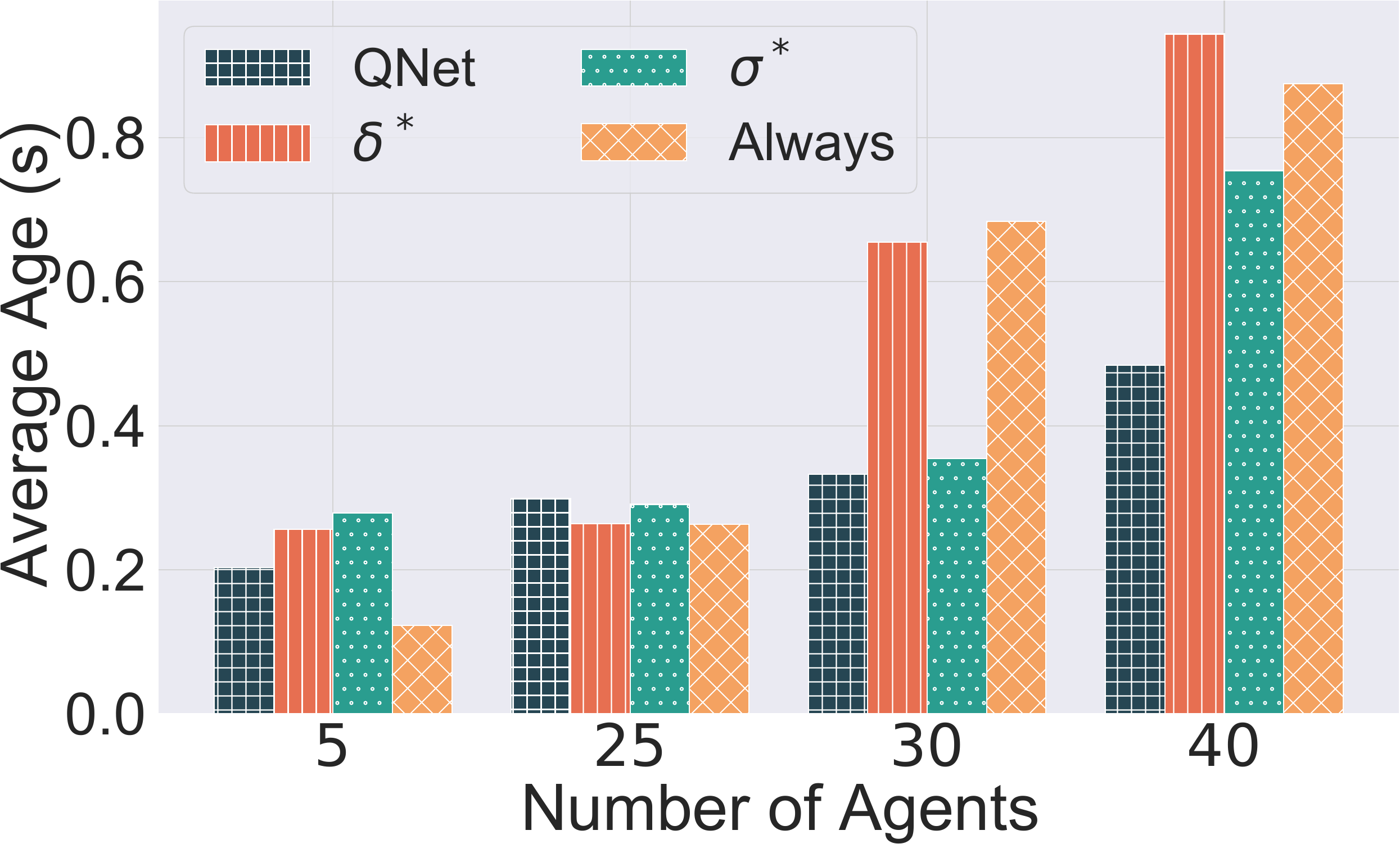}
\label{fig:bar_age_orbit}}
\Description[]{}
\caption{\small Comparison of (a) mean estimation error, (b) query rates, and (c) average age across all agent configurations. $\delta^*$ and $\sigma^*$ represent the best-performing Threshold and Probabilistic Querying Policies, respectively, for the selected number of agents. }
\label{fig:}
\end{center}
\end{figure*}

Table~\ref{tab:orbit_estimation_error} tabulates the mean and standard deviation of estimation error obtained using QNet and the baseline policies, for every choice of number of agents in our experiments. Consider QNet and the Always Query policy. The latter achieves a smaller error than QNet for when we have $5$ agents sharing the network. For larger number of agents QNet does at least as well or better with respect to the mean error. The standard deviation of error for QNet is also either similar to Always Query or smaller. Now consider Threshold Based Querying. The best threshold for a given number of agents has its error boxed in the table. We see that as the number of agents increases, the best performing threshold increases. This satisfies our intuition that an agent must query less frequently as the number of agents sharing the network increases. It also highlights the importance of adapting to network conditions. While there isn't as clear a trend when it comes to Probabilistic Querying, the best variant (boxed) does achieve good performance. 

Further consider Figures~\ref{fig:bar_estimation_orbit},~\ref{fig:bar_age_query}, and~\ref{fig:bar_age_orbit}, that compare the mean estimation error, the query rates, and the obtained average age, respectively. The policies $\delta^*$ and $\sigma^*$ are, respectively, the best Threshold and Probabilistic Querying Policies for the chosen number of agents. Other than for $5$ agents, QNet does either better or almost as well as the baselines. And it does so while keeping the query rates and the average ages that are the smallest or close to the smallest achieved by the best Threshold and Probabilistic Querying Policies. 

As expected, Always Query maintains the same query rate for different numbers of agents sharing the network. Also, as we noted earlier, it is a good choice with regards to keeping error small when it is known that the number of agents sharing the network will be small enough. However, one may pay a penalty of using the network inefficiently. As is seen for when we have $25$ agents, Always Query and QNet have the same mean estimation error, but Always Query has a much larger query rate (about $1.5\times$) than QNet. For larger number of agents, Always Query not only results in worse estimation error but also results in a much greater wastage of network resources. See Figures~\ref{fig:per_orbit},~\ref{fig:rtt_orbit}, and~\ref{fig:retries_exp}. Always Query results in significantly larger PER and retry rates and larger RTT than QNet, when we have more than $25$ agents. The constant rate of querying results in severe network contention.

In summary, QNet allows agents to adapt well to unknown and varied network conditions, which is advantageous in real-world network deployments. Always Query works well for a small number of agents but is expensive in its use of the network. A particular choice (threshold in Threshold Policy and a probability function in Probabilistic Queryinq) is unlikely to work well over a range of network conditions. The need for a wide range of choices further demonstrates the efficacy of QNet in adapting to network conditions.

\section{Transfer To Real Cellular Networks}
\label{sec:cellular_experiments}
We evaluate the transfer of QNet to cellular networks. As in Section~\ref{subsec:orbit_baselines_policies}, we compare QNet with other policies. 

\subsection{Real Cellular Networks Using Traces and Mahimahi}
We test QNet over publicly available real cellular network traces, collected in New York City (NYC), USA~\cite{learning_based_congestion_mahimahi_traces} and Ghent, Belgium~\cite{mahimahi_belgium}. The traces were replayed using the network emulator Mahimahi~\cite{mahimahi_base}. 

The traces from NYC were collected for $4$G LTE networks when walking, driving (in a bus or car), and when the user was stationary. The latter included stationary users in Times Square, a crowded setting characterized by heavy foot traffic. These traces were collected using the tool Saturator~\cite{saturator_app_mahimahi2} to measure packet delivery times at an application running at the user. We have $20$ such traces.

The traces from Ghent were gathered in early $2016$ and spanned transport modes such as foot, bus, tram, train, and car. These network traces recorded bandwidth measurements where users traveled on multiple routes in and around Ghent, actively downloading a large file to capture available bandwidth. An Android application on a Huawei P$8$ Lite logged all relevant data, providing a representative dataset of $4$G network conditions. We have $40$ such traces. 

For both the cities, the LTE networks were real network deployments with the users recording the traces having no control or information about others using the networks. Thus the users were recording available LTE network opportunities to them as they used the networks in the cities.

Each line in a trace file, collected from either city, which serves as an input to Mahimahi, contains a timestamp in milliseconds, representing a packet delivery opportunity, that is a millisecond during which an MTU (Maximum Transmission Unit)-sized packet can be delivered. Accounting is done at the byte-level, and each delivery opportunity represents the ability to deliver $1500$ bytes. The number of repeats of a timestamp indicates the number of MTU-sized packets that can be delivered within the corresponding millisecond. Delivery opportunities are wasted if bytes are unavailable at the instant of an opportunity.

\subsection{Agents, Sources and Edge-Cloud for Cellular Experiments}
We have two agents, each of which obtains measurements from its source, which as in the case of WiFi experiments (Section~\ref{subsec:setup_orbit}), is a SUMO trace. The two agents communicate with an access point (AP) that serves as an edge-cloud node. As a result, we have a cellular uplink between each agent and the AP. Each agent uses its uplink to send its measurements and queries to the AP. The AP uses its downlink to an agent, to respond to the agent's queries. Thus each experiment configuration requires two uplink and two downlink cellular traces. As in Section~\ref{subsec:setup_orbit}, agents send measurements once every $0.1$ second and every agent makes a decision to query or not every $0.1$ second. Also, the measurement packet is $1024$ bytes large.

Broadly, we characterize our experiment as either \emph{stationary} or \emph{not-stationary}. A stationary experiment setup has all its four traces chosen from users that were stationary in Times Square. We have a total of $15$ stationary traces, all from NYC. In a not-stationary setup, the traces are from users that were traveling by foot, bicycle, car, bus, train, or tram. We have a total of $45$ not-stationary traces, of which $5$ are from NYC and the rest are from Ghent. Our characterization is motivated by the differences in the spread of available temporal network throughputs for the two settings, specifically by the RTT experienced by packets that are sent once every $0.1$ second. We will call this RTT as the baseline RTT (bRTT). A temporary reduction in throughput will result in queueing of such packets, increasing the bRTT. For the stationary setups, the bRTT are small and in the range of $0.04 - 0.15$ second. For not-stationary setups the bRTT are in the range of $0.07 - 0.83$ second.


Every experiment, stationary or otherwise, was assigned four traces that were chosen randomly and without replacement from all available traces in the category. We ran $15$ stationary experiments and $32$ not-stationary experiments.

\subsection{Efficacy of Sim-to-Real Transfer}

\begin{figure}[t] 
	\begin{center}
		\includegraphics[height =0.25\linewidth]{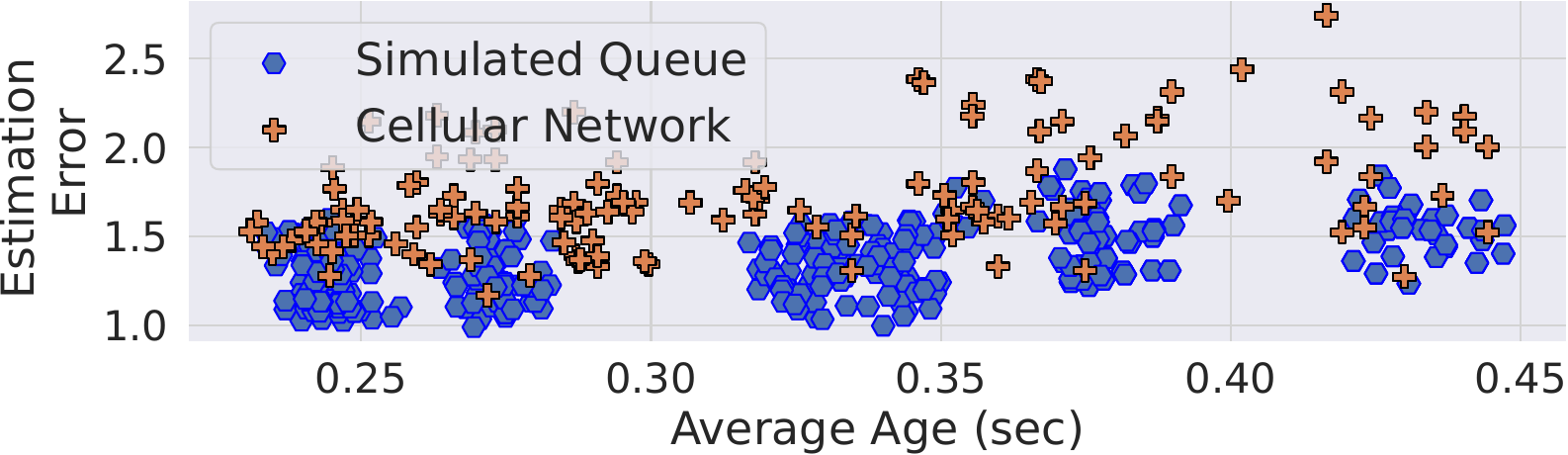}
		\caption{\small A scatter plot of average age and average estimation error for experiments using real Cellular networks overlaid with a scatter obtained using the network simulation model.}
		\label{fig:sim2real_cellular}
	\end{center}
 \Description{}
\end{figure}

Figure~\ref{fig:sim2real_cellular} shows the relationship between average estimation error and average age obtained from simulations and the cellular experiments. Observe that the scatter plots overlap quite well, indicating that the QNet model trained in simulation transferred well to our experiments that used traces from real cellular networks.

Table~\ref{tab:sim2real_cellular_performance} zooms in to show the efficacy of the transfer over different bins of average age. For each bin, we tabulate the average estimation error and the standard deviation. We see that the averages and the standard deviations obtained from experiments on cellular are about the same, with the mean error slightly larger. QNet transfers well to cellular, in a zero-shot manner, in our experiments.

\begin{table}[]
\fontsize{9pt}{9pt}\selectfont
\centering
\caption{\small Mean and standard deviation of estimation error when using QNet, shown for different average age bins}
\label{tab:sim2real_cellular_performance}
\begin{tabular}{@{}cccccc@{}}%
\toprule
\multicolumn{6}{c}{\textbf{Estimation Error (Average $\pm$ Standard Deviation)}} \\ \midrule
\textbf{\begin{tabular}[c]{@{}c@{}}Average Age\\ (sec)\end{tabular}} & \textbf{Simulated} & \multicolumn{1}{c|}{\textbf{Real Cellular}} & \multicolumn{1}{c}{\textbf{\begin{tabular}[c]{@{}c@{}}Average Age\\ (sec)\end{tabular}}} & \multicolumn{1}{c}{\textbf{Simulated}} & \multicolumn{1}{c}{\textbf{Real Cellular}}  \\ \midrule


{[}0.25, 0.3) &1.22 $\pm$ 0.14  & \multicolumn{1}{c|} {1.54 $\pm$ 0.15} & {[}0.3, 0.35) &1.34 $\pm$ 0.12 & \multicolumn{1}{c}  {1.63 $\pm$ 0.18}   \\

{[}0.35, 0.4) & 1.39 $\pm$ 0.12& \multicolumn{1}{c|} {1.80 $\pm$ 0.17} & {[}0.4, 0.45) &  1.48 $\pm$ 0.14 & \multicolumn{1}{c}  {1.84 $\pm$ 0.25}   \\

\bottomrule
\end{tabular}%
\end{table}

\subsection{Evaluation and Comparison with Baselines Over Cellular Networks}
\label{sec:evaluation_cellular}

Table~\ref{tab:table_s_ns} compares the estimation error performance of QNet with Always Query and the best threshold and probabilistic querying policies. For Stationary experiments and not-stationary experiments with bRTT smaller than $0.2$ seconds, we see that all the policies have overlapping error performance. For not-stationary experiments with bRTT greater than $0.2$ seconds does much better than Always Query and Probabilistic Querying. Its performance overlaps with the best Threshold based policy.

\begin{figure*}[]
\begin{center}
\subfloat[\small{Estimation Error}]{\includegraphics[height = 0.17\linewidth] {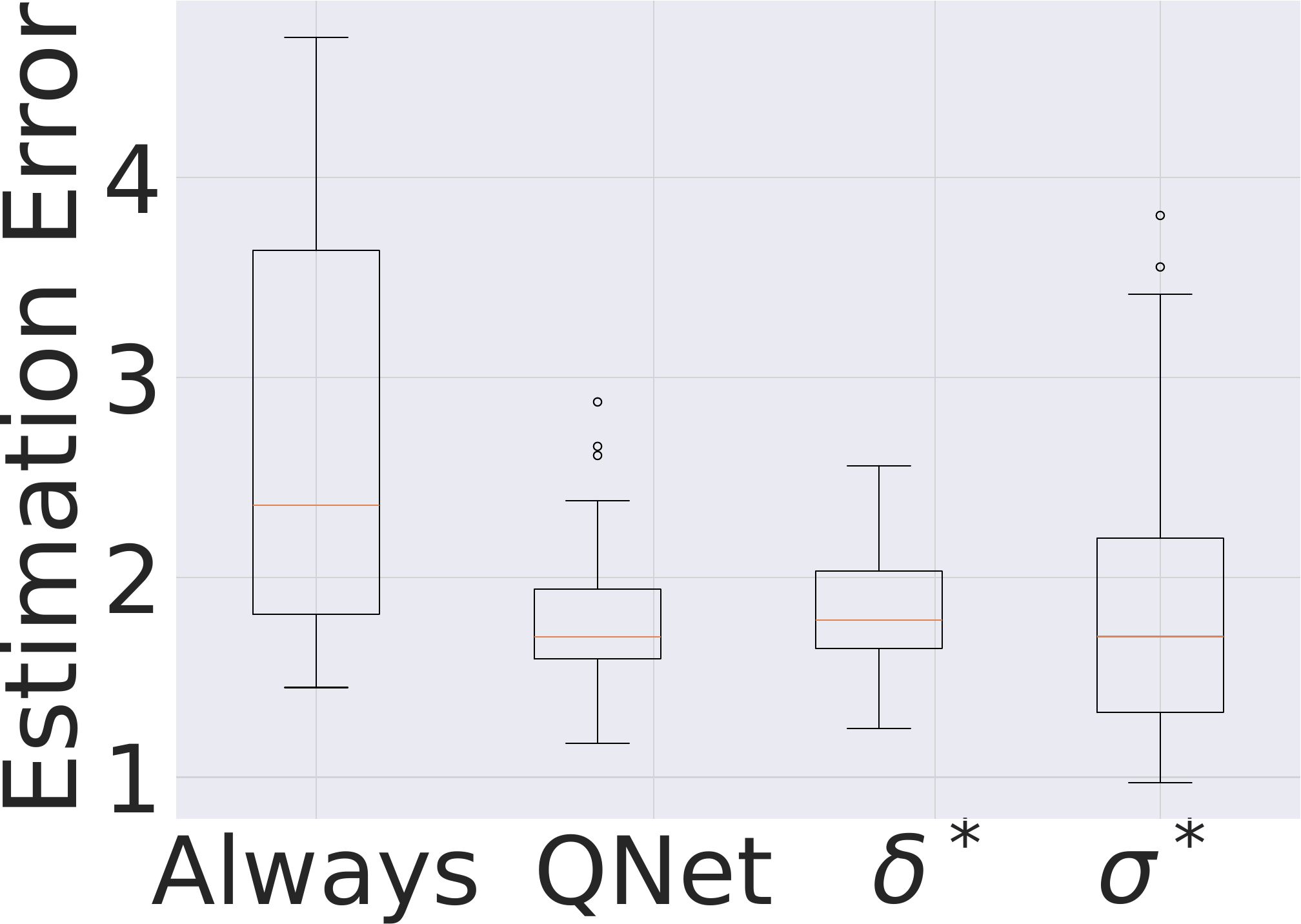}
\label{fig:error_ns_more_0.2}}
\subfloat[\small{Query Rate}]{\includegraphics[height = 0.17\linewidth] {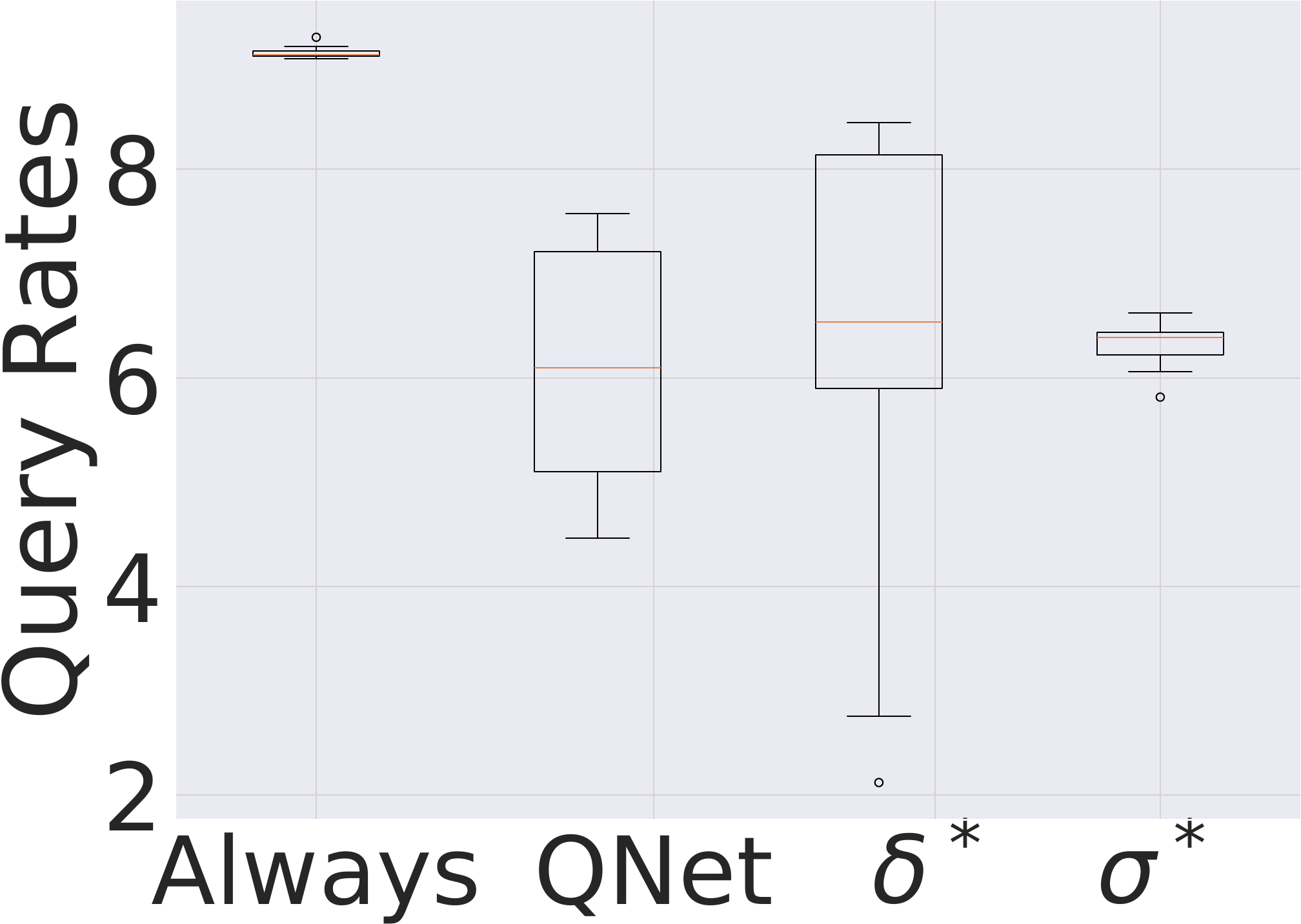}
\label{fig:query_ns_more_0.2}}
\subfloat[\small{Average Age}]{\includegraphics[height = 0.17\linewidth] {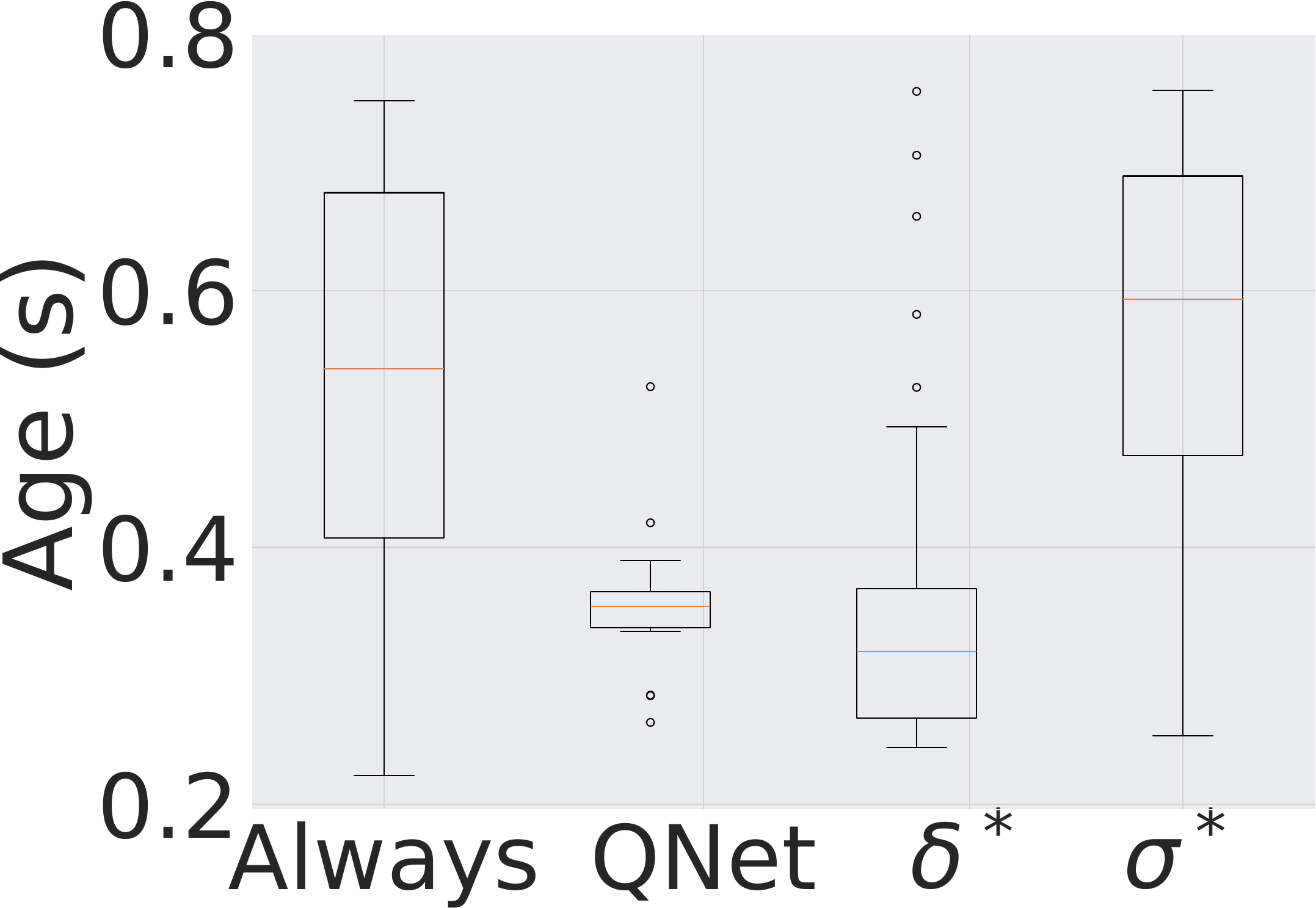}
\label{fig:age_ns_more_0.2}}
\subfloat[Average RTT]{\includegraphics[height = 0.17\linewidth] {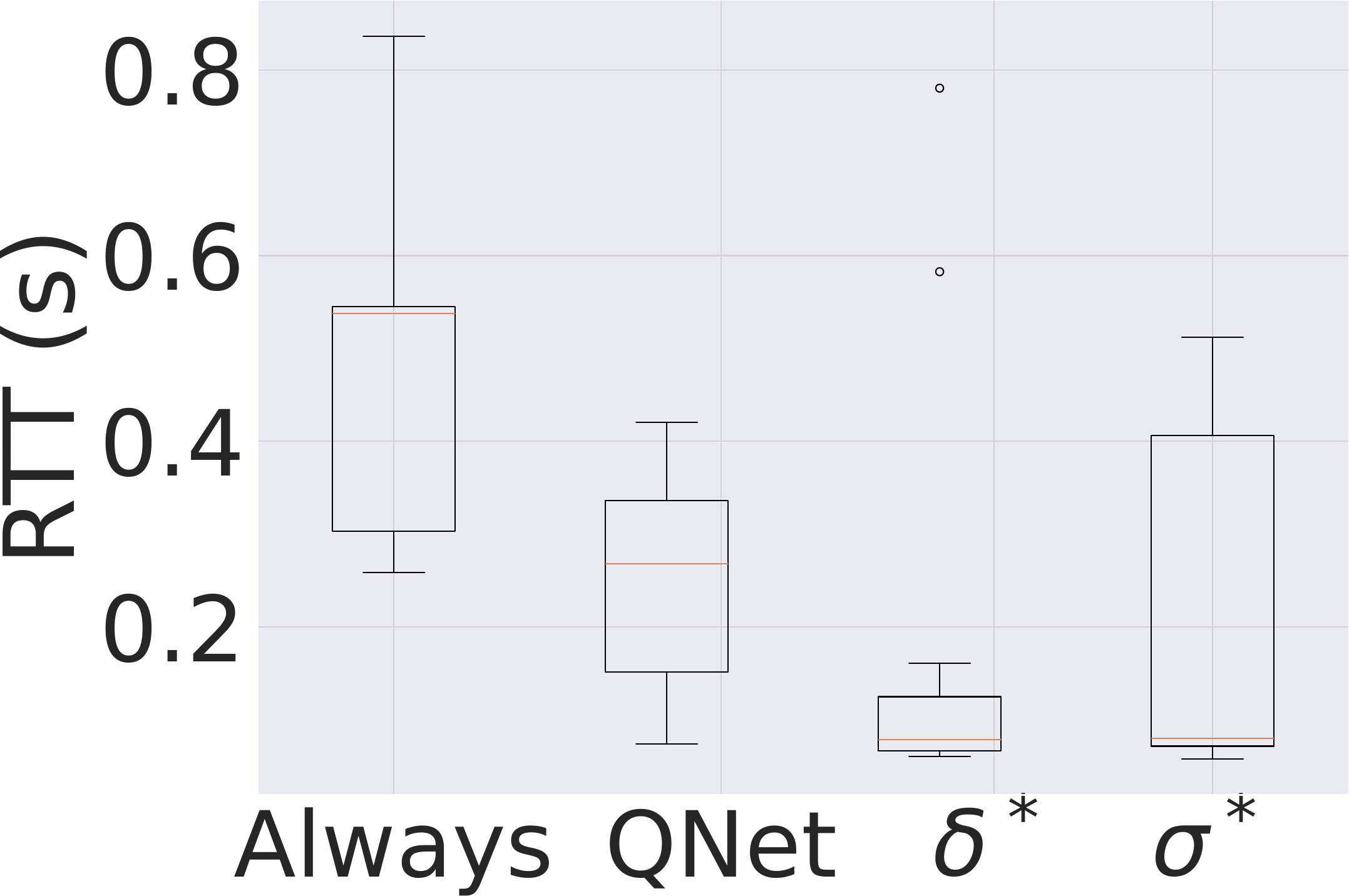}
\label{fig:rtt_ns_more_0}}
\Description[Evaluation of not-stationary experiments with bRTT $\ge0.2$]{}
\caption{Comparison of policies over not-stationary experiments with bRTT$\ge 0.2$ s.}
\label{fig:metrics_non_stationary_ge_0.2}
\end{center}
\end{figure*}

Fig~\ref{fig:metrics_non_stationary_ge_0.2} shows the distribution (box-plot) of estimation error, the query rates, age, and RTT, for the different policies, for not-stationary experiments with bRTT$\ge 0.2$. Note that since Always Query sends a query packet every $0.1$ second, the distribution of RTT for Always Query is also the distribution of bRTT.  Always Query results in a very large spread of estimation errors, with the $75$\textsuperscript{th} percentile about $3.5$ m. QNet and the best threshold based policy have a smaller median error and spread. QNet achieves the low error while using a constrained network, with low available throughputs as evidenced by large bRTT,  efficiently, as is evidenced by its smaller query rate, low age and low RTT. 

\begin{figure*}[]
\begin{center}
\subfloat[\small{Estimation Error}]{\includegraphics[height = 0.17\linewidth] {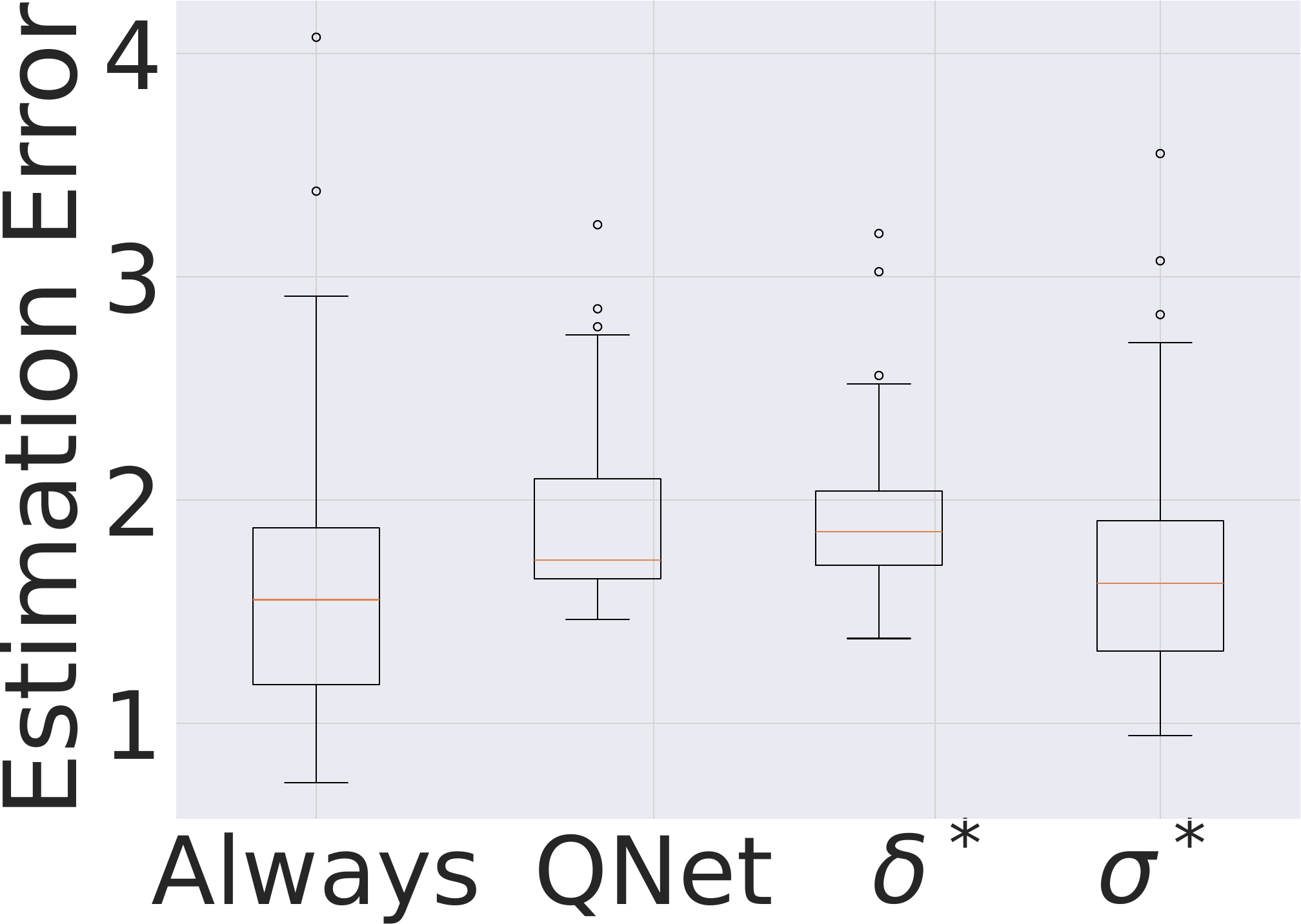}
\label{fig:error_ns_less_0.2}}
\subfloat[\small{Query Rate}]{\includegraphics[height = 0.17\linewidth] {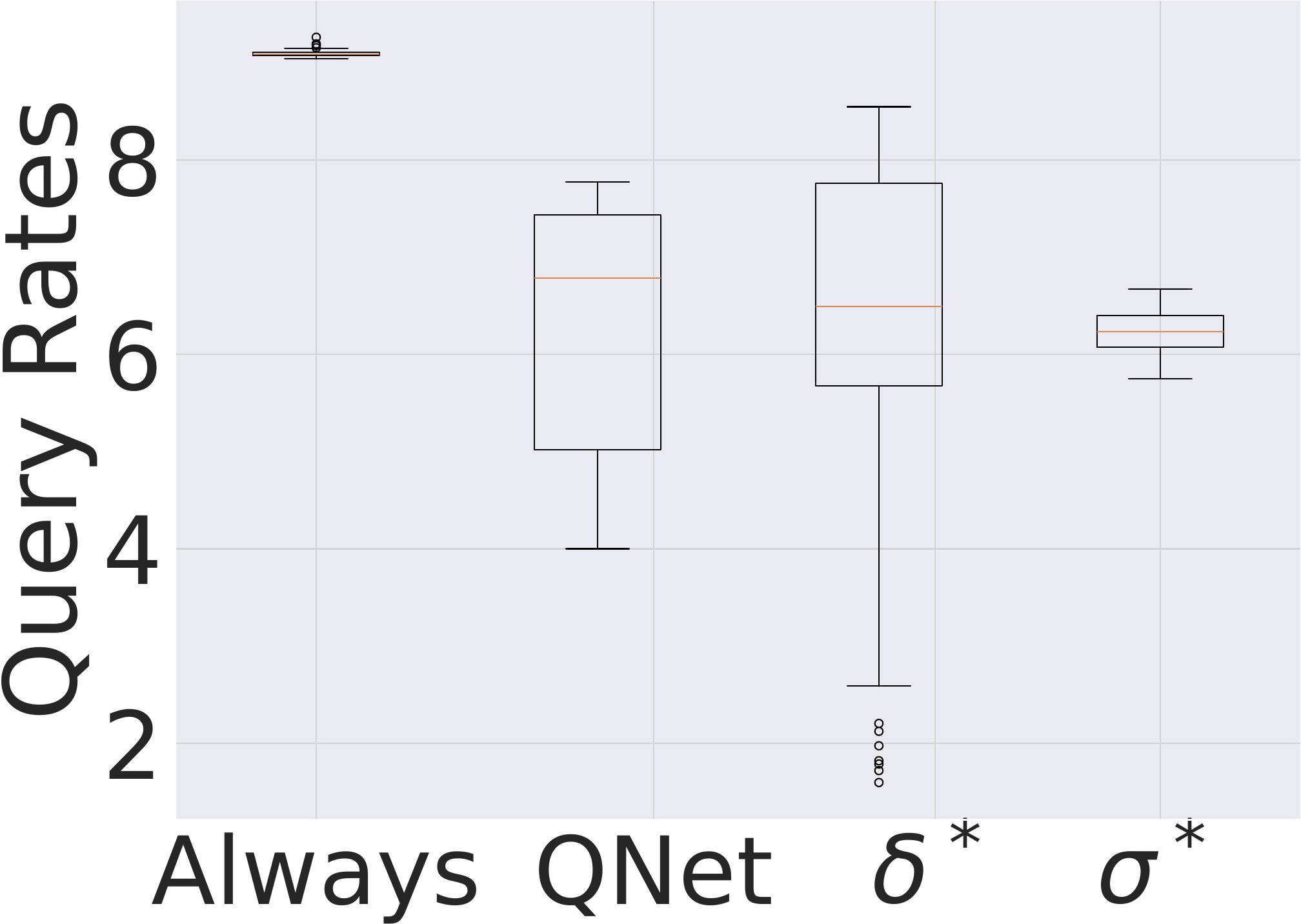}
\label{fig:query_ns_less_0.2}}
\subfloat[\small{Average Age}]{\includegraphics[height = 0.17\linewidth] {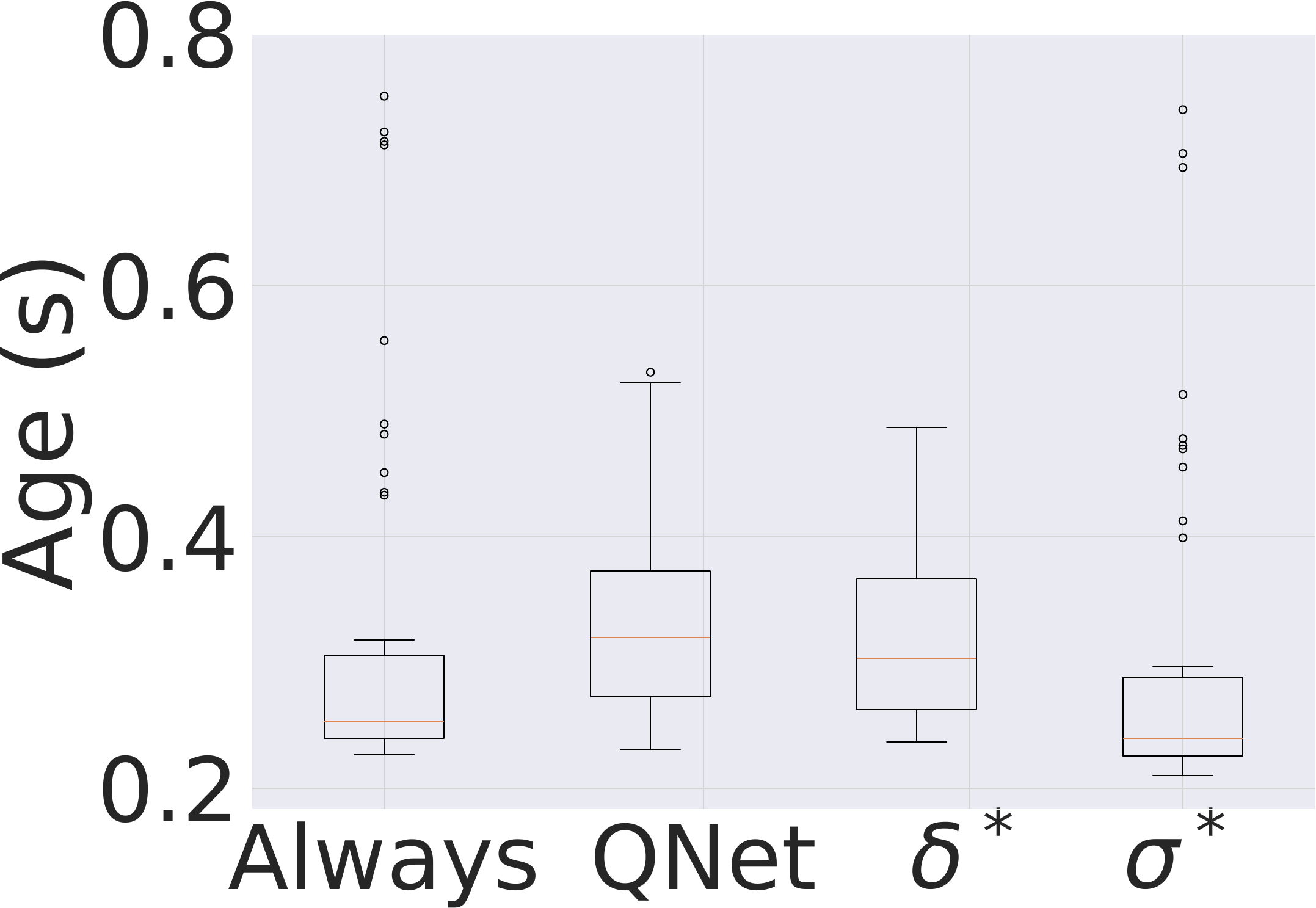}
\label{fig:age_ns_less_0.2}}
\subfloat[\small{Average RTT}]{\includegraphics[height = 0.17\linewidth] {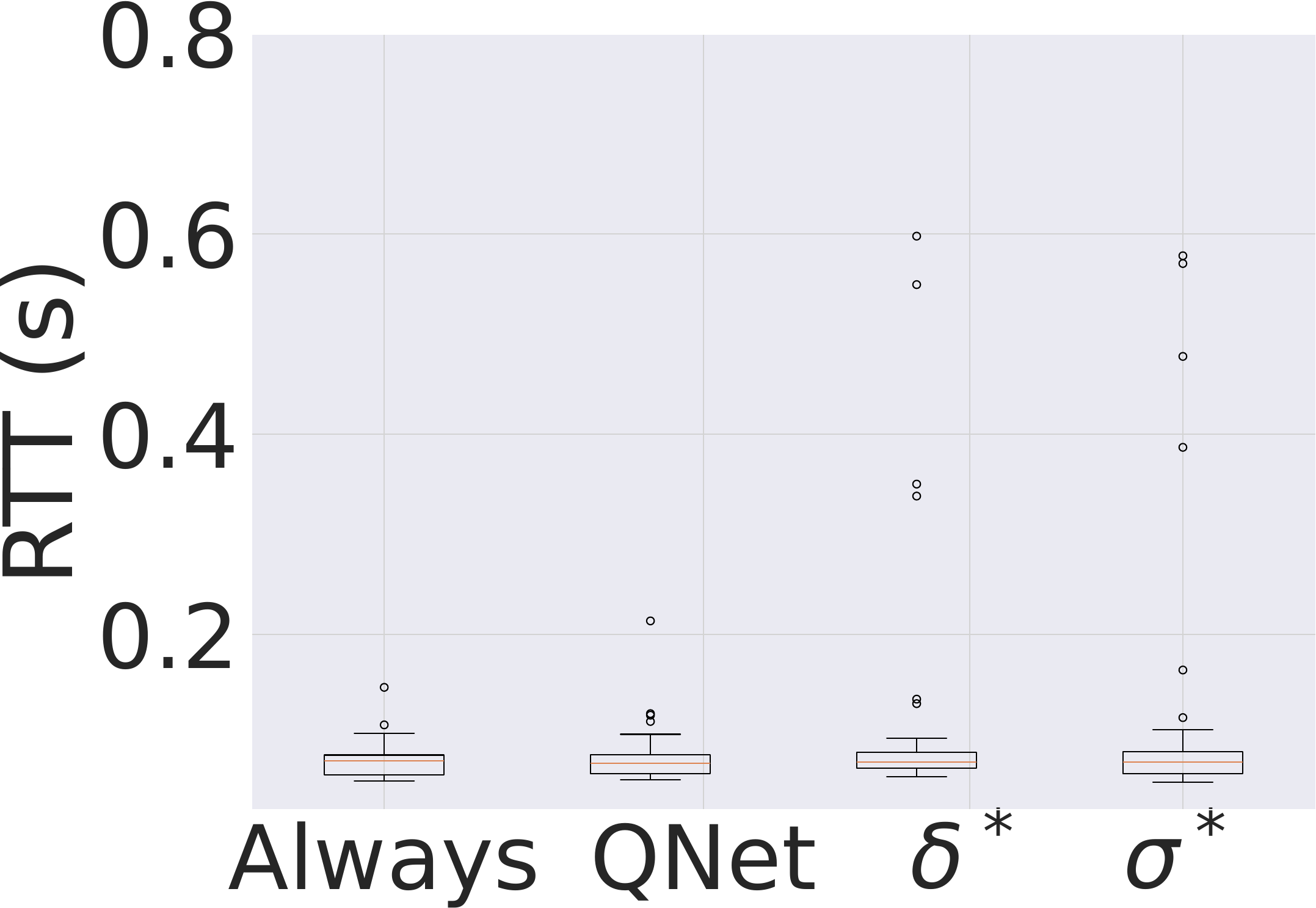}
\label{fig:rtt_ns_less_0.2}}
\Description[Evaluation of not-stationary experiments with bRTT $<0.2$]{}
\caption{Comparison of policies over not-stationary experiments with bRTT$< 0.2$ s.}
\label{fig:metrics_non_stationary_less_0.2}
\end{center}
\end{figure*}

Figure~\ref{fig:metrics_non_stationary_less_0.2} shows the distributions for not-stationary experiments with bRTT$ < 0.2$. The estimation error distributions for the different policies overlap, with Always Query achieving a smaller median error. However, this slight improvement is obtained at a much higher query rate. The higher query rate of Always Query also results in a smaller age, with a RTT similar to the other policies. The good network conditions, as evidenced by the small bRTT, are conducive to the high rate of querying by Always Query and bring small improvements in error and freshness, while not loading the network as evidenced by the small RTT that result from Always Query. Note that unlike in Fig~\ref{fig:metrics_non_stationary_ge_0.2}, the RTT achieved by all policies are as small as the bRTT. Essentially, none of the policies see their packets backlogged. The network conditions are fast enough to allow low delay transfer of measurements and queries from the agents, and responses to queries by the AP.

\begin{figure*}[]
\begin{center}
\subfloat[\small{Estimation Error}]{\includegraphics[height = 0.16\linewidth] {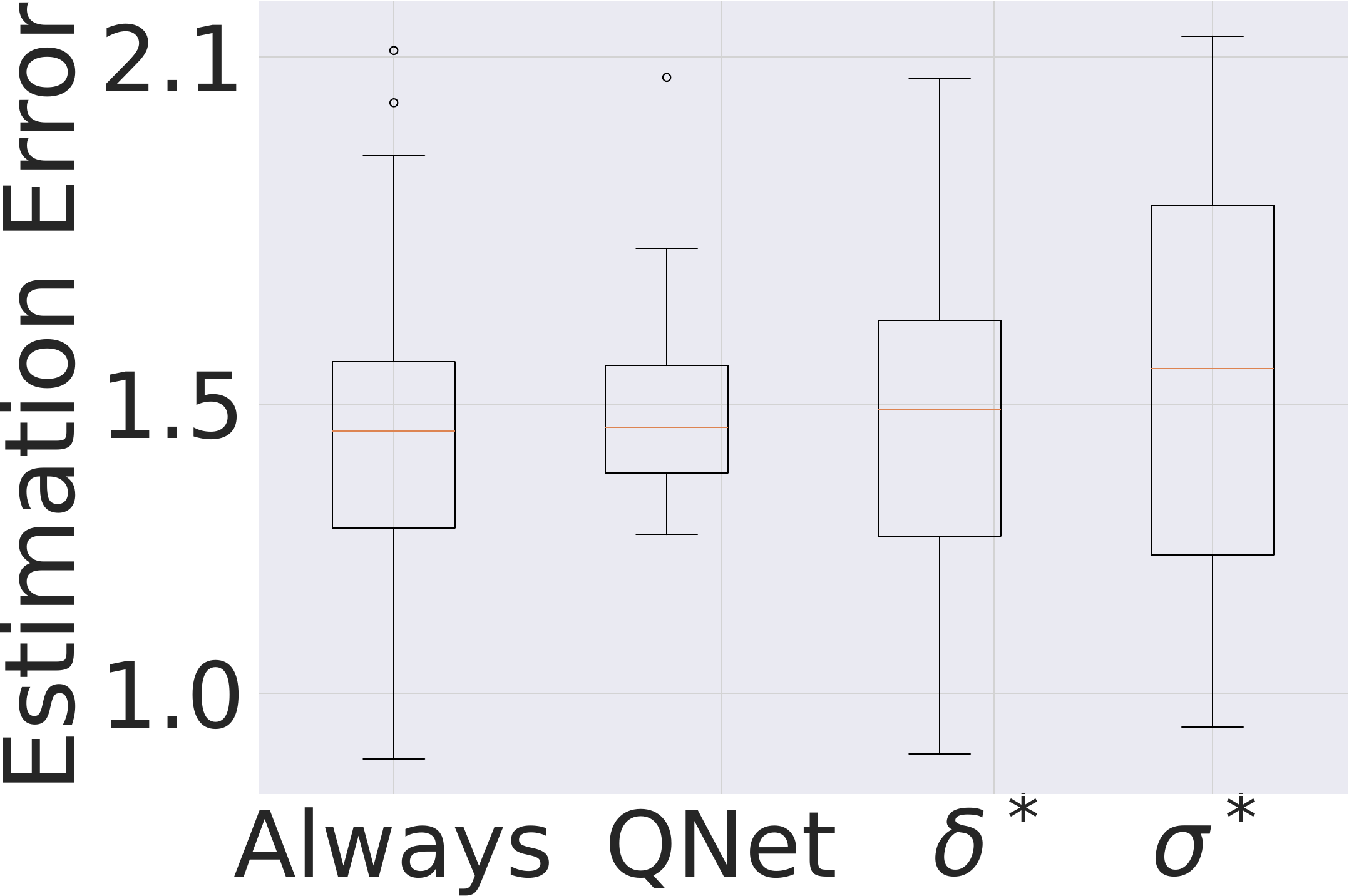}
\label{fig:estimation_stationary}}
\subfloat[\small{Query Rates}]{\includegraphics[height = 0.16\linewidth] {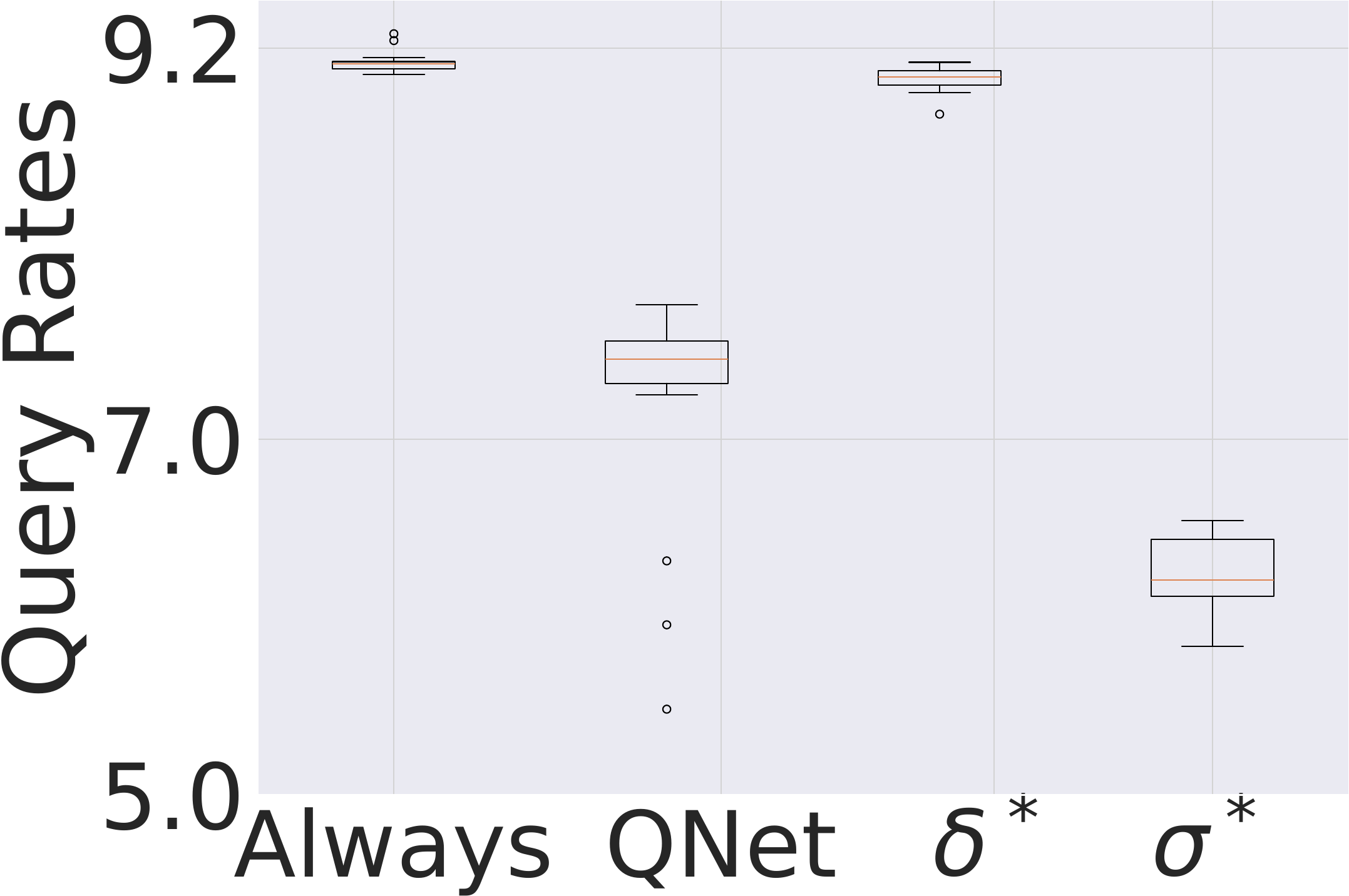}
\label{fig:query_stationary}}
\subfloat[\small{Average Age}]{\includegraphics[height = 0.16\linewidth] {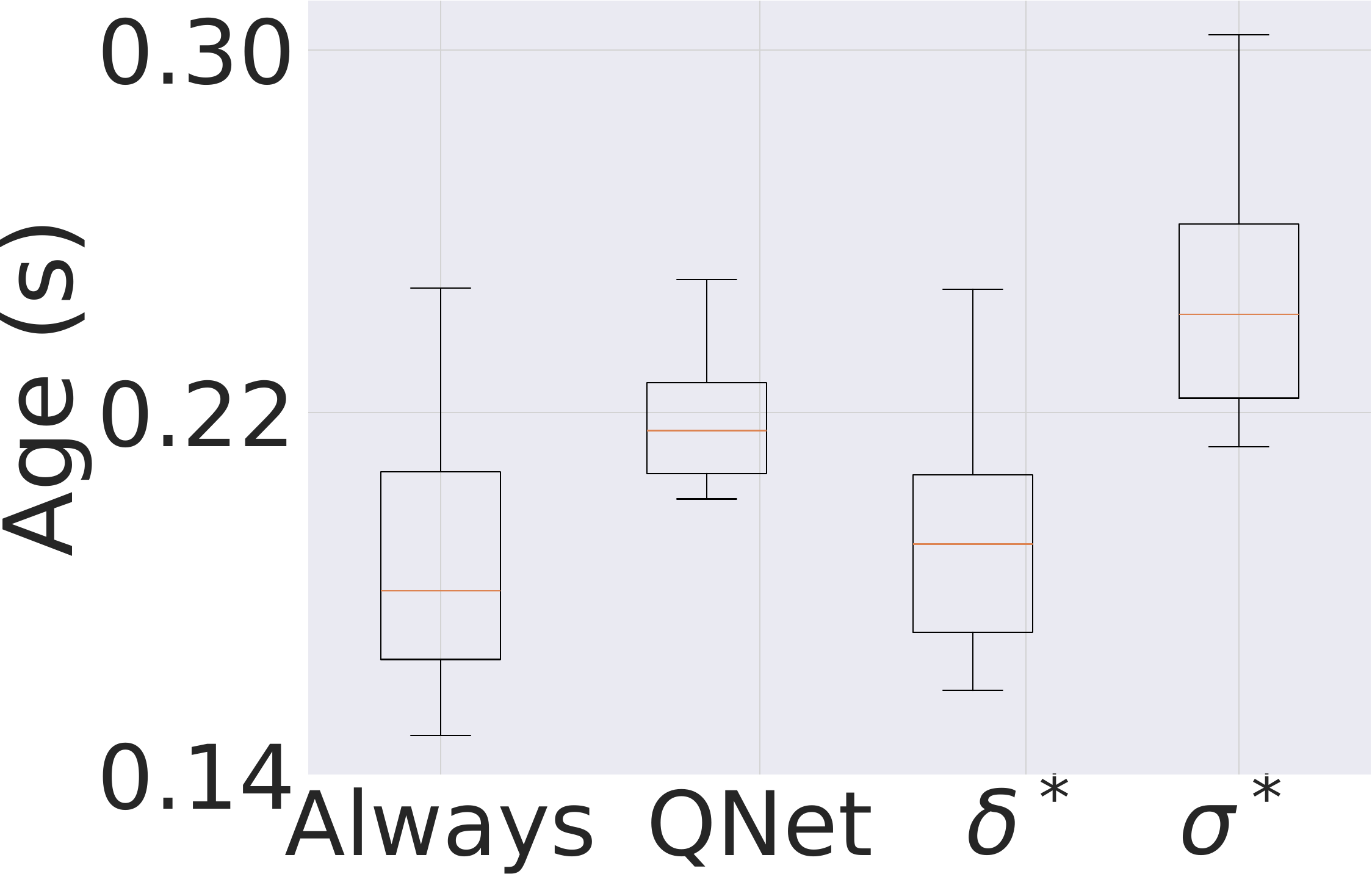}
\label{fig:age_stationary}}
\subfloat[\small{ Average RTT}]{\includegraphics[height = 0.16\linewidth] {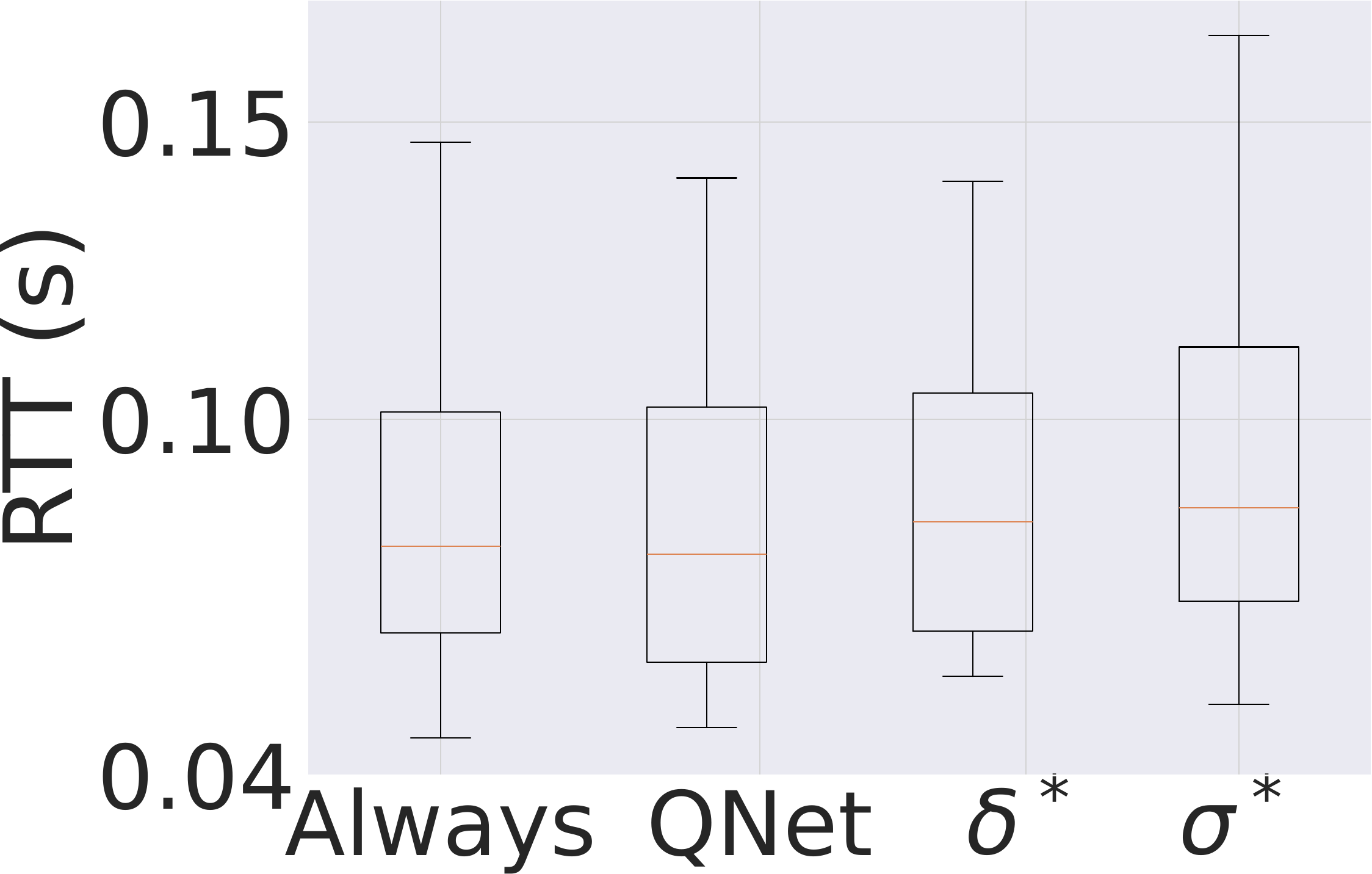}
\label{fig:rtt_stationary}}
\Description[Evaluation of stationary experiments]{}
\caption{Comparison of policies over stationary experiments.}
\label{fig:metrics_stationary}
\end{center}
\end{figure*}

Figure~\ref{fig:metrics_stationary} shows the distributions for stationary experiments. We observe that the median estimation errors are about the same for all policies, with QNet having a smaller spread. QNet's performance comes at a low query rate, while achieving a small enough age and RTT.

\begin{table*}[t]
\fontsize{9pt}{9pt}\selectfont
\centering
\caption{Comparison of mean and standard deviation of performance metrics across all policies, shown separately for stationary and not-stationary experiment setups. For not-stationary, we take a separate look at setups with high ($14$ out of $32$ not-stationary experiments) and low bRTT ($18$ out of $32$).}
\label{tab:table_s_ns}
\resizebox{\textwidth}{!}{%
\begin{tabular}{c|c|c|c|c}
\hline
\textbf{Experiment} & \textbf{Always Query} & \textbf{Threshold Based ($\delta^*$}) & \textbf{Probabilistic Querying ($\sigma^*$}) & \textbf{QNet} \\ \hline

{\textbf{Stationary}}  & \textbf{1.45$\pm$0.33 }& 1.50$\pm$0.30& 1.55$\pm$0.35 &  1.54$\pm$0.25\\

{\textbf{Not-Stationary (bRTT $< 0.2$)}}  & \textbf{1.68$\pm$0.70} & 1.85 $\pm$0.40  &1.72$\pm$0.56  & 1.80$\pm$0.41\\

{\textbf{Not-Stationary (bRTT $\ge 0.2$)}}  & 2.98$\pm$ 1.15
 & 1.92$\pm$0.39  &2.11$\pm$ 0.73  & \textbf{1.82$\pm$ 0.43}\\
\hline
\end{tabular}%
}
\end{table*}

To summarize, QNet does better than the other policies over a constrained network and as well as when the network provides good throughput at low delays (bRTT).

\section{Conclusion}
\label{sec:conclusions}
We proposed a novel deep reinforcement learning model QNet that enables an agent to effectively choose when to query an edge-cloud over a shared wireless network. Further we proposed a simulation-to-real framework for training QNet using only low-fidelity simulations. The framework trains a proposed simulation model for the network and edge-cloud using domain randomization. The simulation model is just single parameter and can be used to generate many episodes in a low-cost manner to train QNet over a wide range of network conditions. We showed via experiments with many agents sharing a real WiFi network and using traces from real cellular networks, that QNet trained in simulation transfers well to real networks. We also compared QNet with other policies. QNet adapts very well to a wide range of network conditions and doesn't require to be trained for any specific network of interest. 

\bibliographystyle{ACM-Reference-Format}
\bibliography{ma_comm,literature_updated,ma_comm_ct_dt}

\end{document}